\documentclass[aps,twocolumn,superscriptaddress]{revtex4}
\usepackage{graphicx}

\newcommand{\bra}[1] {\left\langle #1 \right|}
\newcommand{\ket}[1] {\left| #1 \right\rangle}
\newcommand{\avg}[1] {\left\langle #1 \right\rangle}

\begin{document}

\title{The information about the state of a qubit gained by a
weakly coupled detector}

\author{S. Ashhab}
\affiliation{Advanced Science Institute, The Institute of Physical
and Chemical Research (RIKEN), Wako-shi, Saitama 351-0198, Japan}
\affiliation{Physics Department, Michigan Center for Theoretical
Physics, The University of Michigan, Ann Arbor, Michigan
48109-1040, USA}

\author{J. Q. You}
\affiliation{Advanced Science Institute, The Institute of Physical
and Chemical Research (RIKEN), Wako-shi, Saitama 351-0198, Japan}
\affiliation{Department of Physics and Surface Physics Laboratory
(National Key Laboratory), Fudan University, Shanghai 200433,
China}

\author{Franco Nori}
\affiliation{Advanced Science Institute, The Institute of Physical
and Chemical Research (RIKEN), Wako-shi, Saitama 351-0198, Japan}
\affiliation{Physics Department, Michigan Center for Theoretical
Physics, The University of Michigan, Ann Arbor, Michigan
48109-1040, USA}

\date{\today}



\begin{abstract}
We analyze the information that one can learn about the state of a
quantum two-level system, i.e.~a qubit, when probed weakly by a
nearby detector. In particular, we focus on the case when the
qubit Hamiltonian and the qubit's operator being probed by the
detector do not commute. Because the qubit's state keeps evolving
while being probed and because the measurement data is mixed with
a detector-related background noise, one might expect the detector
to fail in this case. We show, however, that under suitable
conditions and by proper analysis of the measurement data useful
information about the state of the qubit can be extracted. It
turns out that the measurement basis is stochastically determined
every time the experiment is repeated. We analyze in detail the
probability distributions that govern the choice of measurement
bases. We also analyze the information acquisition rate and show
that it is largely unaffected by the apparent conflict between the
measurement and intrinsic qubit dynamics. We discuss the relation
between our analysis and the stochastic master equation that
describes the evolution of the qubit's state under the influence
of measurement and decoherence. In particular, we write down a
stochastic equation that encompasses the usual stochastic master
equation for the evolution of the qubit's density matrix and
additionally contains the measurement information that can be
extracted from the observed signal.
\end{abstract}

\maketitle

\section{Introduction}

Quantum measurement is a subject of importance both as a central
part of quantum theory \cite{Zurek} and for its relevance to
quantum information theory \cite{Nielsen}. At the most basic
level, the measurement is commonly thought of as a sudden
projective operation that leaves the measured system in an
eigenstate of the probed operator. It should be kept in mind,
however, that the measurement process is a physical process that
follows from a physical interaction between the measured system
and the measuring device. As a result, any realistic measurement
is carried out over a finite duration. One can therefore have the
situation where the measurement process occurs concurrently with
other physical processes. This is the situation that we consider
in this paper.

Although the analysis and results of this paper should be valid
for a variety of different physical settings, we find it useful to
formulate the analysis in terms of a specific setup. In this
paper, we use the example of a double-quantum-dot charge qubit
that is probed by a quantum point contact (QPC)
\cite{Gurvitz,Korotkov,Makhlin,Goan,Pilgram,Jordan,Johansson,Ouyang,Reuther,Clerk,Romito,Kurotani,QuantumCapacitance,Kreisbeck}.
The principle of the measurement is the fact that the current in
the QPC is sensitive to the charge state of the qubit. By
measuring the current passing through the QPC, one can infer the
state of the qubit. One limitation that arises in practical
situations is that, in order to minimize the effects of the
detector on the qubit at times when no measurement is performed
(e.g.~during the implementation of quantum gates), the
qubit-detector coupling is set to a value that is small compared
to the qubit's energy scale. As a result one must deal with some
form of weak-measurement regime. This type of weak,
charge-sensitive readout works well when the qubit is biased such
that the charge states are eigenstates of the Hamiltonian and
therefore do not mix during the measurement. In this case one can
allow the detector to probe the qubit for as long as is needed to
obtain a high signal-to-noise ratio, without having to worry about
any intrinsic qubit dynamics.

In contrast to the simple situation described above, when the
detector weakly probes the charge state of the qubit while the
Hamiltonian induces mixing dynamics between charge states, it
becomes unclear how to interpret a given measurement signal. Given
the fact that the signal typically contains a large amount of
detector-related noise and the measurement causes unavoidable
relaxation in the qubit, it might seem that this type of
measurement cannot be used to determine the initial state of the
qubit, i.e.~at the time that the experimenter decides to perform
the measurement. Indeed, there have been a number of studies
analyzing the measurement-induced decoherence and the evolution of
the qubit's state in this situation
\cite{Gurvitz,Korotkov,Makhlin,Goan,Pilgram,Jordan,Kurotani}, but
not the question of how to take the measurement data and extract
from it information about the initial state of the qubit. This
question is a key issue for qubit-state readout and is the main
subject of this paper.

We shall show below that high-fidelity measurement information can
be extracted from the measurement data, provided that additional
decoherence mechanisms are weak and the readout signal can be
monitored at a sufficiently short timescale. It turns out that not
only the measurement result, but also the measurement basis, is
determined stochastically in this case. In spite of the
uncontrollability of the measurement basis, the measurement
results contain meaningful information about the initial state of
the qubit. In order to demonstrate this statement, we show how
these results can be used to perform quantum state tomography on
the qubit. These results show that under suitable conditions and
by proper analysis of the measurement data useful information
about the state of the qubit can be extracted from this data.

This paper is organized as follows: In Sec.~II we introduce the
theoretical model for describing a charge qubit probed by a QPC.
In Sec.~III we formulate the theoretical framework for the qubit
evolution conditioned on the QPC's output signal and how
measurement information can be extracted from this signal. We also
introduce an equation that supplements the stochastic master
equation for purposes of interpreting the measured QPC signal. In
Sec.~IV we present results of numerical calculations that display
a number of interesting results in this system, most importantly
the stochastic nature of the measurement basis and the
near-independence of the measurement fidelity from the relation
between the charge basis and energy eigenbasis. In Sec.~V we
discuss how decoherence can be introduced to the problem and how
one should interpret the measurement data in the presence of
decoherence. We analyze one simple example that shows how the
measurement data recorded at later times is less `valuable' than
that recorded at earlier times. Concluding remarks are given in
Sec.~VI. Some details of the calculations are given in the
appendices.

\section{Model}

\begin{figure}[h]
\includegraphics[width=8.0cm]{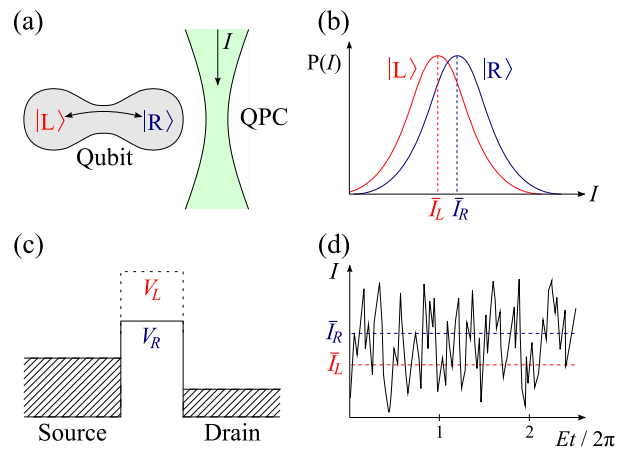}
\caption{(color online) Schematic diagrams of (a) a charge qubit
with charge states $\ket{L}$ and $\ket{R}$ measured by a quantum
point contact (QPC), (b) the probability distributions of the
possible QPC current values for the two charge states of the
qubit, (c) the filled energy levels of the QPC's electrodes and
the energy barriers for tunneling between the electrodes for the
two charge states of the qubit and (d) a typical QPC current
signal. The finite widths of the probability distributions in (b)
are a result of the finite measurement time. When the distance
between the two center points, $|\overline{I}_R-\overline{I}_L|$,
is much smaller than the widths of the distributions, the QPC
performs a weak measurement on the qubit in the short interval
under consideration (i.e., the time interval for which these
probability distributions are relevant). The fast fluctuations in
(d) convey the assumption that these fluctuations have a much
shorter timescale than the qubit's precession period. In plotting
(b-d) we have assumed that $\overline{I}_R>\overline{I}_L$, which
would be the case if the qubit is defined by an extra positive
charge (e.g.~a hole) tunneling between the two wells. Taking the
opposite case, i.e.~$\overline{I}_L>\overline{I}_R$, would only
result in a few sign changes in the analysis below. The main
results are not affected by this choice.}
\end{figure}

We consider a system composed of a charge qubit capacitively
coupled to a QPC, as illustrated in Fig.~1. The qubit can be
viewed as a system where a charged particle is trapped in a
double-well potential and can occupy, and tunnel between, the
localized ground states of the two wells. We shall denote these
states by $\ket{L}$ and $\ket{R}$.

During the measurement a voltage is applied to the QPC, and a
current flows through it. Except for the discussion given in
Sec.~V, we shall assume that the QPC does not induce any qubit
decoherence except that associated with the measurement-induced
projection. Before analyzing the general case, we first consider
the case where the qubit Hamiltonian is diagonal in the charge
basis. In this case, there is no mechanism by which the states
$\ket{L}$ and $\ket{R}$ mix during the system dynamics. As a
result, if the qubit is initially in the state $\ket{L}$, the
long-time-averaged QPC current will be given by $\overline{I}_L$,
and the qubit will remain in the state $\ket{L}$. A similar
statement applies to the state $\ket{R}$ of the qubit, with
corresponding QPC current $\overline{I}_R$. The QPC current
therefore serves as an indicator of the qubit's state in the
charge basis $\{\ket{L},\ket{R}\}$, as long as the qubit
Hamiltonian does not mix the states of this basis.

On any finite timescale, there will be fluctuations in the QPC
current, and the observed value might deviate from
$\overline{I}_L$ or $\overline{I}_R$. The longer the period over
which the averaging is made, the smaller the fluctuations. One can
therefore define a measurement timescale that determines how long
one needs to wait in order to distinguish between the states
$\ket{L}$ and $\ket{R}$. The relation between this timescale and
the qubit's Hamiltonian-induced precession period separates two
measurement regimes: strong and weak qubit-detector coupling. As
mentioned above, this separation is irrelevant when the qubit
Hamiltonian is diagonal in the charge basis.

For the remainder of this paper, we analyze the general case where
the qubit Hamiltonian is not necessarily diagonal in the charge
basis. The Hamiltonian of the combined qubit+QPC system can be
expressed as
\begin{equation}
\hat{H} = \hat{H}_{\rm q} + \hat{H}_{\rm QPC} + \hat{H}_{\rm int}.
\end{equation}
In describing the qubit, we shall use the eigenbasis of the qubit
Hamiltonian $\hat{H}_{\rm q}$. Thus $\hat{H}_{\rm q}$ can be
expressed as
\begin{equation}
\hat{H}_{\rm q} = \left( \begin{array}{cc}
-\frac{E}{2} & 0 \\
0 & \frac{E}{2} \\
\end{array}
\right) = - \frac{E}{2} \hat{\sigma}_z,
\end{equation}
where $E$ is the energy splitting between the qubit's two energy
levels, and $\hat{\sigma}_z$ is the $z$-axis Pauli matrix. We
shall express the ground and excited states of the Hamiltonian as
$\ket{0}$ and $\ket{1}$, respectively. The electric dipole moment
operator $\ket{R}\bra{R} - \ket{L}\bra{L}$ can be expressed as
$\hat{\sigma}_{\bf n}$, where $\bf n$ is a vector that represents
the direction of the charge basis relative to the energy
eigenbasis:
\begin{eqnarray}
\hat{\sigma}_{\bf n} & = & \ket{R}\bra{R} - \ket{L}\bra{L}
\nonumber \\
& = & \cos\beta \; \hat{\sigma}_z + \sin\beta \; \hat{\sigma}_x,
\end{eqnarray}
where $\beta$ represents the angle between the charge basis and
the energy eigenbasis. The states of the charge basis (i.e., the
eigenstates of $\hat{\sigma}_{\bf n}$) can be expressed as
\begin{eqnarray}
\ket{R} & = & \cos\frac{\beta}{2} \ket{0} + \sin\frac{\beta}{2}
\ket{1} \nonumber
\\
\ket{L} & = & \sin\frac{\beta}{2} \ket{0} - \cos\frac{\beta}{2}
\ket{1}.
\end{eqnarray}
The QPC Hamiltonian $\hat{H}_{\rm QPC}$ can be expressed as
\begin{eqnarray}
\hat{H}_{\rm QPC} & = & \sum_m \epsilon_{S,m} \hat{a}_m^{\dagger}
\hat{a}_m + \sum_n \epsilon_{D,n} \hat{b}_n^{\dagger} \hat{b}_n
\nonumber \\ & & + \sum_{\scriptsize \begin{array}{c}\{m,n\}\in \\
{\rm tunneling} \\ {\rm channels} \end{array}} J_{m,n} \left(
\hat{b}_n^{\dagger} \hat{a}_m + \hat{a}_m^{\dagger} \hat{b}_n
\right),
\end{eqnarray}
where the operators $\hat{a}_m$ and $\hat{b}_n$, respectively,
annihilate electrons in the source and drain electrodes of the
QPC, $\hat{a}_m^{\dagger}$ and $\hat{b}_n^{\dagger}$ are their
hermitian conjugates, $\epsilon_{S,m}$ and $\epsilon_{D,n}$ are
the energies of single-particle quantum states in the source and
drain electrodes, and $J_{m,n}$ represents tunneling matrix
elements between the two electrodes. We take the qubit-QPC
interaction Hamiltonian to be of the form \cite{Gurvitz}
\begin{equation}
\hat{H}_{\rm int} = \hat{\sigma}_{\bf n} \sum_{\scriptsize
\begin{array}{c}\{m,n\}\in \\ {\rm tunneling} \\ {\rm channels}
\end{array}} \delta \! J_{m,n} \left( \hat{b}_n^{\dagger} \hat{a}_m
+ \hat{a}_m^{\dagger} \hat{b}_n \right),
\end{equation}
such that the tunneling matrix elements (and therefore the QPC
current) depend on the state of the qubit in the charge basis. The
QPC and its operation are illustrated in Figs.~1(b-d). When $J$
and $\delta \! J$ are independent of $m$ and $n$, the microscopic
degrees of freedom of the QPC are not affected by, and therefore
do not carry any information about, the state of the qubit. This
information is carried only by the total number of electrons that
have tunneled through the QPC. In the remainder of this paper, we
shall assume that this is the case. We close this section by
mentioning that the above description of the microscopic structure
of the QPC is not needed for the analysis below, provided one
accepts the description of the QPC operation explained in Sec.~III
as a phenomenological model of the detector. Indeed, the analysis
below is not restricted to the case of a charge qubit probed by a
QPC.

\section{Measurement- and Hamiltonian-induced dynamics}

In this section we introduce the theoretical framework for
analyzing the qubit state evolution and interpreting the observed
measurement data.

\subsection{Short-time evolution}

We start our analysis by considering a short time interval between
times $t$ and $t+\delta t$. For definiteness, we assume that
during this time interval a large number of electrons tunnel
through the QPC, such that it is natural to define a QPC current
$I(t)$ during this short interval (The current would be defined as
the amount of charge that has passed through the QPC divided by
$\delta t$; The small-current case will be discussed in
Sec.~IV.F). We also assume that a weak-coupling regime exists for
a properly chosen value of $\delta t$, which means that the
QPC-current probability distributions (for the states $\ket{L}$
and $\ket{R}$) are broad and almost completely overlap, as shown
in Fig.~1(b). The coherence time of the QPC and the time
resolution of the QPC's output signal are both assumed to be much
shorter than the precession period of the qubit. As a result, for
a properly chosen value of $\delta t$ one can treat the different
time intervals as independent measurement processes, with each
measurement being made in the charge basis.

As explained in Appendix A, the QPC-current probability
distribution for the qubit state $\ket{L}$ can be expressed as
\begin{equation}
P_L[I,\delta \! I,\delta t] = \sqrt{\frac{2\delta t \left( \delta
\! I \right)^2}{\pi \tau_m \left( \Delta \overline{I} \right)^2}}
\exp \left\{ - \frac{2\delta t \left(
I-\overline{I}_L\right)^2}{\tau_m
\left(\Delta\overline{I}\right)^2} \right\},
\end{equation}
and a similar expression can be derived for the qubit state
$\ket{R}$. Here we have defined a number of useful quantities:
$\delta \! I$ is the size of a finite interval of QPC currents
that we identify with a single value. One could say that with this
definition we are turning the probability distributions in
Fig.~1(b) into histograms with discrete possible values for the
current $I$. This definition will be useful for both the analytic
and numerical calculations below. The characteristic measurement
time $\tau_m$ defines the timescale needed for the QPC output to
distinguish between the states $\ket{L}$ and $\ket{R}$, and
$\Delta\overline{I}=\overline{I}_R-\overline{I}_L$, where for
definiteness we take $\overline{I}_R>\overline{I}_L$.

We can now construct matrices that describe the qubit-state
evolution depending on the observed QPC current $I(t)$. When a
given value of $I(t)$ is observed in the QPC, the density matrix
$\rho_{\rm q}(t)$ describing the quantum state of the qubit is
projected (possibly partially) according to the observed value.
This projection of the qubit's state is described by a $2 \times
2$ matrix (or propagator) that we shall call
$\hat{U}_M[I(t),\delta \! I,\delta t]$:
\begin{widetext}
\begin{equation}
\rho_{\rm q}(t+\delta t) = \frac{1}{{\rm Tr}
\left\{\hat{U}_M^{\dagger}[I(t),\delta \! I,\delta
t]\hat{U}_M[I(t),\delta \! I,\delta t] \rho_{\rm q}(t) \right\}}
\hat{U}_M[I(t),\delta \! I,\delta t] \rho_{\rm q}(t)
\hat{U}_M^{\dagger}[I(t),\delta \! I,\delta t].
\label{Eq:Density_matrix_projection}
\end{equation}
Measurement theory says that $\hat{U}_M[I(t),\delta \! I,\delta
t]$ obeys the equation \cite{Nielsen,Jacobs}
\begin{equation}
\hat{U}_M^{\dagger}[I(t),\delta \! I,\delta t]
\hat{U}_M[I(t),\delta \! I,\delta t] = P_L[I(t),\delta \! I,\delta
t] \ket{L}\bra{L} + P_R[I(t),\delta \! I,\delta t] \ket{R}\bra{R}.
\label{Eq:UMdaggerUM}
\end{equation}
Since we are keeping only the fundamentally necessary terms in the
projection dynamics, we ignore any unitary-transformation
component in $\hat{U}_M[I(t),\delta \! I,\delta t]$ that does not
affect Eq.~(\ref{Eq:UMdaggerUM}). Any such component can in
principle be determined experimentally, and it can then be
incorporated into the analysis straightforwardly; Such a component
could result if, for example, the different current states of the
QPC have different charge distributions that act as an effective
gate voltage on the charge qubit. Assuming a large overlap between
the QPC-current probability distributions, we find that the matrix
$\hat{U}_M[I(t),\delta \! I,\delta t]$ can be expressed in the
form
\begin{eqnarray}
\hat{U}_M[I(t),\delta \! I,\delta t] & = & \sqrt{P_L[I(t),\delta
\! I,\delta t]} \ket{L}\bra{L} + \sqrt{P_R[I(t),\delta \! I,\delta
t]} \ket{R}\bra{R} \nonumber \\
& = & \left( \frac{2\delta t \left( \delta \! I \right)^2}{\pi
\tau_m \left( \Delta \overline{I} \right)^2} \right)^{1/4} \left[
\exp \left\{ - \frac{\delta t \left[ I(t)-\overline{I}+\Delta
\overline{I}/2\right]^2}{\tau_m \left(\Delta\overline{I}\right)^2}
\right\} \ket{L}\bra{L} + \exp \left\{ - \frac{\delta t \left[
I(t)-\overline{I}-\Delta \overline{I}/2\right]^2}{\tau_m
\left(\Delta\overline{I}\right)^2} \right\} \ket{R}\bra{R} \right]
\nonumber \\
& \propto & \exp \left\{ -\frac{\delta t \left[ I(t) -
\overline{I} \right]}{\tau_m \Delta\overline{I}} \right\}
\ket{L}\bra{L} + \exp\left\{ \frac{\delta t \left[ I(t) -
\overline{I} \right]}{\tau_m \Delta\overline{I}} \right\}
\ket{R}\bra{R}
\nonumber \\
& \approx & 1 + \frac{\delta t \left[ I(t) - \overline{I}
\right]}{\tau_m \Delta\overline{I}} \left( \ket{R}\bra{R} -
\ket{L}\bra{L} \right) \nonumber \\
& = & 1 + \frac{\delta t \left[ I(t) - \overline{I}
\right]}{\tau_m \Delta\overline{I}} \hat{\sigma}_{\bf n},
\label{Eq:UM}
\end{eqnarray}
where we have defined
$\overline{I}=(\overline{I}_L+\overline{I}_R)/2$.

In addition to the measurement-induced evolution described by the
matrix $\hat{U}_M$ of Eq.~(\ref{Eq:UM}), the qubit Hamiltonian
induces a unitary evolution of the qubit's state. This
contribution to the qubit-state evolution (over the time interval
from $t$ to $t+\delta t$) is described by the matrix
\begin{eqnarray}
\hat{U}_H [\delta t] = \exp\left\{-i \hat{H}_{\rm q} \delta t
\right\} \approx 1 + i \frac{E}{2} \delta t \hat{\sigma}_z.
\end{eqnarray}

The matrices $\hat{U}_M[I(t),\delta \! I,\delta t]$ and $\hat{U}_H
[\delta t]$ can now be combined to give the total evolution matrix
\begin{eqnarray}
\hat{U}[I(t),\delta \! I,\delta t] & \approx &
\hat{U}_M[I(t),\delta \! I,\delta t] \times \hat{U}_H [\delta t]
\nonumber \\
& \propto & 1 + \frac{\delta t \left[ I(t) - \overline{I}
\right]}{\tau_m \Delta\overline{I}} \hat{\sigma}_{\bf n} + i
\frac{E}{2} \delta t \hat{\sigma}_z.
\label{Eq:U_combined}
\end{eqnarray}
Note that the high-resolution requirement in the QPC-current
signal, which is one of our assumptions mentioned earlier in this
section, enters in obtaining the above equation: both
$\hat{U}_M[I,\delta \! I,\delta t]$ and $\hat{U}_H [\delta t]$ are
approximately proportional to the unit matrix, with lowest-order
corrections of order $\delta t$. Therefore they commute to first
order in $\delta t$, and they can be treated as commuting
operators when $E \delta t \ll 1$, i.e.~when they both induce
small changes to the quantum state of the qubit.

When a given QPC output signal $I(t)$ [starting from the initial
time $t=0$ until the final time $t_f$] is observed, one can take
the corresponding short-time evolution matrices explained above
and use them to construct the total evolution matrix $\hat{U}_{\rm
Total}[I(t:0\rightarrow t_f),\delta \! I,\delta t]$. This matrix
describes the evolution of the qubit state from $t=0$ until
$t=t_f$, given that the current signal $I(t)$ was observed (with
discretization parameters $\delta \! I$ and $\delta t$). Using the
unit matrix as the total evolution matrix for $t=0$, we find that
\begin{equation}
\hat{U}_{\rm Total}[I(t:0\rightarrow t_f),\delta \! I,\delta t] =
\hat{U}[I(t_f-\delta t),\delta \! I,\delta t] \times \cdots \times
\hat{U}[I(0),\delta \! I,\delta t].
\label{Eq:U_Total}
\end{equation}
In Sec. III.B we shall explain how measurement information can be
extracted from the matrix $\hat{U}_{\rm Total}$.

\subsection{Extracting information from the matrix
$\hat{U}_{\rm Total}[I(t:0\rightarrow t_f), \delta \! I,\delta
t]$}

The evolution of the qubit state is obtained by taking into
account a long sequence of weak measurements and
Hamiltonian-induced unitary transformations. This long sequence of
events, however, is equivalent to a simple scenario in which a
single measurement is made on the qubit and the state of the qubit
is rotated after the measurement. In order to obtain a
quantitative description of this alternative interpretation of the
measurement data, one takes the $2 \times 2$ matrix $\hat{U}_{\rm
Total}[I(t:0\rightarrow t_f),\delta \! I,\delta t]$ and divides it
through to a polar decomposition into two parts, a measurement
matrix $\hat{U}_{\rm Meas}[I(t:0\rightarrow t_f),\delta \!
I,\delta t]$ followed by a unitary transformation $\hat{U}_{\rm
Rot}[I(t:0\rightarrow t_f),\delta \! I,\delta t]$:
\begin{equation}
\hat{U}_{\rm Total}[I(t:0\rightarrow t_f),\delta \! I,\delta t] =
\hat{U}_{\rm Rot}[I(t:0\rightarrow t_f),\delta \! I,\delta t]
\times \hat{U}_{\rm Meas}[I(t:0\rightarrow t_f),\delta \! I,\delta
t].
\end{equation}
\end{widetext}
The matrix $\hat{U}_{\rm Meas}[I(t:0\rightarrow t_f),\delta \!
I,\delta t]$ has the form
\begin{equation}
\hat{U}_{\rm Meas}[I(t:0\rightarrow t_f),\delta \! I,\delta t] =
\sqrt{P_1} \ket{\psi_1}\bra{\psi_1} + \sqrt{P_2}
\ket{\psi_2}\bra{\psi_2},
\end{equation}
where $\ket{\psi_1}$ and $\ket{\psi_2}$ are two orthogonal states
(the indices are assigned such that $P_1 \geq P_2$). In order to
characterize the measurement matrix in terms of the information it
provides about the state of the qubit, it is useful to consider
the case where one initially has no information about this state.
The qubit therefore starts in the maximally mixed state
\begin{equation}
\rho_{\rm max. \ mixed} = \left(
\begin{array}{cc}
\frac{1}{2} & 0 \\
0 & \frac{1}{2}
\end{array}
\right).
\end{equation}
After a measurement that produces the outcome corresponding to the
matrix $\hat{U}_{\rm Meas}[I(t:0\rightarrow t_f),\delta \!
I,\delta t]$ (and ignoring the presence of the rotation matrix
$\hat{U}_{\rm Rot}[I(t:0\rightarrow t_f),\delta \! I,\delta t]$
for purposes of this argument), the qubit ends up in a state
described by the density matrix
\begin{equation}
\rho_{\rm after \ Meas.} = \frac{P_1}{P_1+P_2}
\ket{\psi_1}\bra{\psi_1} + \frac{P_2}{P_1+P_2}
\ket{\psi_2}\bra{\psi_2}.
\end{equation}
This density matrix describes a statistical mixture of the states
$\ket{\psi_1}$ and $\ket{\psi_2}$, with a higher probability for
the former (except for the rare cases where $P_1=P_2$). One can
therefore say that, as a result of the measurement, one now knows
that the qubit is more likely to be in the state $\ket{\psi_1}$
than in the state $\ket{\psi_2}$, in contrast to the complete lack
of information at the initial time. The states $\ket{\psi_1}$ and
$\ket{\psi_2}$ therefore represent the measurement basis that
corresponds to the output signal $I(t)$. The measurement fidelity
is calculated as follows \cite{AshhabMeasurement}: Let us assume
that one is given a qubit that is in an unknown state, either
$\ket{\psi_1}$ or $\ket{\psi_2}$. The parameters $P_i$ are the
probabilities that the outcome defined by $I(t)$, $\delta \! I$
and $\delta t$ is obtained given that the qubit was initially in
the state $\ket{\psi_i}$. Upon observing this outcome, if one
wants to make a guess about the state of the qubit, one would
maximize the probability of making a correct guess by choosing the
state $\ket{\psi_1}$. The difference between the probability of
actually finding the state $\ket{\psi_1}$ and that of finding the
orthogonal state $\ket{\psi_2}$ is given by
\begin{equation}
F = \left|\frac{P_1-P_2}{P_1+P_2}\right|.
\end{equation}
This expression can therefore naturally be interpreted as the
measurement fidelity.

It is worth pausing here to comment on the issue of the
independence of the measurement basis from the measurement
outcome. One typically thinks of the measurement basis as being
determined by the experimenter when designing the experimental
setup and the measurement result (i.e., $+1$ or $-1$ along the
measurement axis) being determined stochastically according to the
rules of quantum mechanics based on the measurement basis used in
the experiment. The situation considered in this paper, however,
cannot be described using this simple picture. Here we have a
two-state quantum system and a measurement device that produces
one of many possible outcomes, not necessarily all providing
information in the same basis. In this case, the picture of the
measurement basis being independent of the outcome is not valid in
general. For example, it is possible in a general setting for a
certain state $\ket{\psi}$ to be a possible result of the data
analysis presented above, but not the state orthogonal to it. Such
a situation arises in the problem analyzed in
Ref.~\cite{AshhabMeasurement}, and it complicates the
interpretation of the measurement data. Fortunately, because in
this paper we are dealing with two symmetric probability
distributions [see Fig.~1(b)], we find that for every possible QPC
signal there is an `opposite' or `flipped' signal, which can be
obtained by taking the mirror image about the central line in
Fig.~1(d) [this central line is defined by
$\overline{I}=(\overline{I}_L+\overline{I}_R)/2$]. It is not
difficult to verify that if a given signal corresponds to a
certain measured state $\ket{\psi}$, the `opposite' signal will
correspond to the orthogonal state with the same values of $P_1$
and $P_2$ (see Appendix B for the full derivation). An important
consequence of the above statement is that the total probability
of finding one of two opposite signals is independent of the
initial state of the qubit. Hence, the probability of obtaining a
certain measurement basis is independent of the initial state of
the qubit. This fact leads naturally to the conceptual
interpretation that the measurement basis is determined according
to some stochastic process, and the measurement result along that
basis is made after the basis has been determined.

To summarize, the QPC's output signal can be used to derive the
matrix $\hat{U}_{\rm Total}[I(t:0\rightarrow t_f),\delta \!
I,\delta t]$. This matrix can then be used to determine the
measurement basis, the measurement result (i.e., $\pm 1$ along the
measurement axis), the fidelity (or in other words, the degree of
certainty about the obtained measurement result) and the
post-measurement rotation. Note that all the analysis above is
performed independently of the initial state of the qubit. If one
knows the initial state, one can use the evolution matrices in
order to determine the probabilities of different outcomes, as
well as the final state of the qubit after the measurement. As we
shall show below, when the measurement fidelity approaches one,
the final state can be determined even without any knowledge about
the initial state.

It is worth mentioning here that the analysis presented above can
be applied not only to qubits, but also to systems with
higher-dimensional Hilbert spaces. The main complication in the
case of higher dimensions is that the description of the results
becomes more intricate and less transparent. In this context one
can contrast the simple visualization of any qubit density matrix
as a point in the so-called Bloch sphere, whereas there is no
visually simple parametrization for a general density matrix of a
three-state system.

\subsection{Incorporating the matrix
$\hat{U}_{\rm Total}[I(t:0\rightarrow t_f), \delta \! I,\delta t]$
into the formalism of stochastic master equations}

Equation (\ref{Eq:Density_matrix_projection}) describes the
projection of the qubit's density matrix conditioned on the
observation of the QPC output $I(t)$. That equation treats the
evolution as occurring in discrete steps. Constructing a master
equation means taking Eq.~(\ref{Eq:Density_matrix_projection}) to
first order in $\delta t$, and then taking the limit $\delta
t\rightarrow 0$ in order to obtain the corresponding
continuous-time differential equation \cite{Jacobs}. One tricky
point in carrying out this procedure is the fact that the
stochastic quantity $I(t)-\overline{I}$ has fluctuations of size
$1/\sqrt{\delta t}$:
\begin{equation}
\left\langle \left(I(t)-\overline{I}\right)^2\right\rangle =
\frac{\tau_m \left( \Delta\overline{I} \right)^2}{4\delta t}.
\end{equation}
As a result, when keeping terms only to first order in $\delta t$,
one must also keep terms containing $\left(\delta t\right)^2
\times \left(I(t)-\overline{I}\right)^2$, etc. The quantity
$I(t)-\overline{I}$ can be expressed as its expectation value at
any give point in time, which is given by
$\Delta\overline{I}\langle\hat{\sigma}_{\bf n}\rangle/2$, plus a
fluctuation. It is customary to write the stochastic master
equation not in terms of the current fluctuation, but rather in
terms of a rescaled quantity, $\delta W$, whose standard deviation
is equal to $\sqrt{\delta t}$. It then follows that $\delta W$
should be defined using the relation
\begin{equation}
I(t) = \overline{I} + \frac{\Delta\overline{I}}{2}
\langle\hat{\sigma}_{\bf n}\rangle + \frac{\Delta\overline{I}
\sqrt{\tau_m}}{2 \delta t} \delta W.
\end{equation}

Substituting Eq.~(\ref{Eq:U_combined}) into
Eq.~(\ref{Eq:Density_matrix_projection}) [ignoring the subscript
$M$ here] and keeping terms to first order in $\delta t$, we find
that
\begin{widetext}
\begin{equation}
\rho_{\rm q}(t+\delta t) = \rho_{\rm q} - i \delta t \hat{H}_{\rm
q} \rho_{\rm q} + i \delta t \rho_{\rm q} \hat{H}_{\rm q} +
\frac{\delta W}{2\sqrt{\tau_m}} \left( \hat{\sigma}_{\bf n}
\rho_{\rm q} + \rho_{\rm q} \hat{\sigma}_{\bf n} - 2 \left\langle
\hat{\sigma}_{\bf n} \right\rangle \rho_{\rm q} \right) +
\frac{\delta t}{4\tau_m} \left( \hat{\sigma}_{\bf n} \rho_{\rm q}
\hat{\sigma}_{\bf n} - \rho_{\rm q} \right),
\end{equation}
where, for compactness, we have not expressed the time dependence
of $\rho_{\rm q}$ explicitly on the right-hand side [In other
words $\rho_{\rm q}$ should be interpreted as $\rho_{\rm q}(t)$].
This equation can now be rewritten in the form
\begin{equation}
\dot{\rho}_{\rm q} = - i \left[ \hat{H}_{\rm q} , \rho_{\rm q}
\right] + \frac{\xi(t)}{2\sqrt{\tau_m}} \left( \hat{\sigma}_{\bf
n} \rho_{\rm q} + \rho_{\rm q} \hat{\sigma}_{\bf n} - 2
\left\langle \hat{\sigma}_{\bf n} \right\rangle \rho_{\rm q}
\right) + \frac{1}{4\tau_m} \left( \hat{\sigma}_{\bf n} \rho_{\rm
q} \hat{\sigma}_{\bf n} - \rho_{\rm q} \right).
\label{Eq:SME}
\end{equation}
\end{widetext}
where $\xi(t)$ is obtained by using the rescaling $\xi(t)=\delta
W/\delta t$ and taking the limit $\delta t\rightarrow 0$:
\begin{eqnarray}
\left\langle \xi(t) \right\rangle & = & 0 \nonumber \\
\left\langle \xi(t) \xi(t') \right\rangle & = & \delta(t-t'),
\end{eqnarray}
where $\delta(t-t')$ is the Dirac delta function. Equation
(\ref{Eq:SME}) is the stochastic master equation for the system
under consideration.

The question now is how the matrix $\hat{U}_{\rm
Total}[I(t:0\rightarrow t_f), \delta \! I,\delta t]$ fits into the
picture of the stochastic master equation. One can derive a
stochastic equation for the matrix $\hat{U}_{\rm
Total}[I(t:0\rightarrow t_f), \delta \! I,\delta t]$ by
considering the change caused by a matrix of the form given in
Eq.~(\ref{Eq:U_combined}); note that the evolution of the matrix
$\hat{U}_{\rm Total}[I(t:0\rightarrow t_f), \delta \! I,\delta t]$
is governed by Eq.~(\ref{Eq:U_Total}). One problem that arises
here is that the matrix in the first line of
Eq.~(\ref{Eq:U_combined}) tends to zero when $\delta t \rightarrow
0$. Taking this limit therefore leads to ill-defined quantities.
In order to avoid these problems, we deal with a re-normalized
version of the matrix $\hat{U}[I(t),\delta \! I, \delta t]$, as
given in the second line of Eq.~(\ref{Eq:U_combined}); it should
be noted here that normalization of the matrix $\hat{U}_{\rm
Total}$ is irrelevant for purposes of extracting the measurement
basis and fidelity. It is now straightforward to see that an
unnormalized version of the matrix $\hat{U}_{\rm Total}$ obeys the
equation
\begin{equation}
\frac{d \hat{U}_{\rm Total}}{dt} = \left( \left[
\frac{\left\langle\hat{\sigma}_{\bf n}\right\rangle}{2\tau_m} +
\frac{\xi(t)}{2\sqrt{\tau_m}} \right] \hat{\sigma}_{\bf n} - i
\hat{H}_{\rm q} \hat{\sigma}_z \right) \hat{U}_{\rm Total}.
\label{Eq:SME_for_U}
\end{equation}
This equation supplements the stochastic master equation for
purposes of interpreting the QPC's output signal in terms of a
measurement result. In this context, one should note that what
Eq.~(\ref{Eq:SME}) gives is the qubit's density matrix as a
function of time. When $\beta\neq 0$ and the charge states mix
during the measurement, the density matrix at later times can
become completely different from that at the start of the
measurement process. Assuming that the experimenter does not have
any information about the qubit's state at $t=0$, he would be able
to use Eq.~(\ref{Eq:SME}) to determine the qubit's state at later
times, but he would not be able to cast the results in terms of
information that he has learned about the qubit's state at $t=0$.
The matrix $\hat{U}_{\rm Total}$, through the decomposition
explained in Sec.~III.B contains the information needed to make
such a statement about the measurement result. In fact, one can
say that, when taken in combination with the initial qubit density
matrix, Eq.~(\ref{Eq:SME_for_U}) replaces Eq.~(\ref{Eq:SME}): in
addition to containing the measurement information, the matrix
$\hat{U}_{\rm Total}$ can be used to calculate the density matrix
at any time.

\section{Numerical results and discussion}

We now present the results of our numerical calculations. Typical
parameters of the numerical calculations are as follows: The
discrete steps are taken to be $\delta t=0.1\pi/E$, such that one
period of the Hamiltonian-induced coherent oscillations would be
divided into 20 steps. The QPC current probability distributions
are discretized into 100 possible current values: each
distribution is Gaussian with a standard deviation $\sigma$ of 10
steps (thus there are about 50 possible values for the QPC current
with non-negligible occurrence probability). The distance between
the two center points of the current probability distributions
$\Delta \overline{I}$ determines the qubit-QPC coupling strength.
This parameter is varied when analyzing the effect of the
qubit-QPC coupling strength on the behaviour of the system. The
number of time steps in a single calculation was different for
different calculations. However, this number was always large
enough that the final qubit state reached perfect purity up to
less than one part in $10^5$, ensuring that none of our numerical
results were sensitive to the exact choice of this parameter. Each
measurement procedure is repeated up to $10^4$ times in order to
obtain accurate statistical averages. We have also used parameters
different from the typical ones given above and found that the
results are unaffected by the specific choice of these parameters.

A convenient parameter for purposes of characterizing the
qubit-QPC coupling strength is $E\tau_m/(2\pi)$, where $\tau_m$ is
the timescale needed to obtain sufficient QPC signal to read out
the state of the qubit (for the time being one can think of this
definition as applying to the case when $\beta=0$; but see below).
As the standard deviation of the QPC-output effective probability
distribution scales as $\sigma/\sqrt{N}$ when the measurement step
is repeated $N$ times, the measurement time $\tau_m$ can be
naturally defined as the product of the time step $\delta t$ and
the value of $N$ at which $2\sigma/\sqrt{N}=\Delta \! I$. $\tau_m$
is therefore given by
\begin{equation}
\tau_m = \frac{4\sigma^2 \delta t}{\Delta I^2}.
\end{equation}
We shall use the parameter $E\tau_m/(2\pi)$ for quantifying the
qubit-QPC coupling strength, and we shall use $\tau_m$ as a
characteristic measurement timescale when presenting our results
below.

\subsection{Stochastically determined measurement basis}

\begin{figure}[h]
\includegraphics[width=9.0cm]{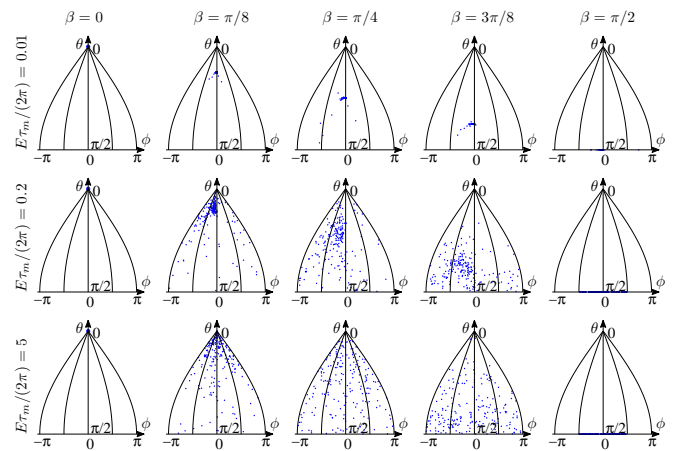}
\caption{(color online) The spherical coordinates $\theta$ and
$\phi$ defining the stochastically determined measurement bases
obtained in simulations of the experiment under consideration
(note that, since the measurement basis is defined by an axis
extending in two opposite directions, only the upper hemisphere is
needed; any measurement axis will either lie in the plane of the
equator or have one end above the equator). Each figure contains
200 points. In each row, the qubit-QPC coupling strength is kept
fixed and $\beta$ is varied: $\beta=0$, $\pi/8$, $\pi/4$, $3\pi/8$
and $\pi/2$. In each column $\beta$ is kept fixed and the
qubit-QPC coupling strength is varied. In the top row,
$E\tau_m/(2\pi)=0.01$, which is deep in the strong-coupling
regime; The measurement basis is very close to the charge basis in
about 99\% of the runs. In the middle row, $E\tau_m/(2\pi)=0.2$,
which can be identified as the intermediate-coupling regime; The
measurement basis deviates substantially from the charge basis in
about one half of the runs. In the bottom row $E\tau_m/(2\pi)=5$,
which is in the weak-coupling regime; The measurement bases are,
in general, spread over the entire hemisphere. In generating this
figure, the initial state was taken to be the maximally mixed
state (note that the results are independent of the initial
state).}
\end{figure}

First, in Fig.~2 we show the spherical coordinates $\theta$ and
$\phi$ of a large number of (stochastically determined)
measurement bases for different qubit-detector coupling strengths
and different values of $\beta$. In the strong-coupling regime
(top row in Fig.~2), the parameter $E\tau_m/(2\pi)$ is much
smaller than one, and the measurement is completed before any
Hamiltonian-induced dynamics can occur. As a result, the
measurement basis is always the charge basis, i.e.~the natural
measurement basis for the detector under consideration [note here
that the charge basis is characterized by the direction
$(\theta,\phi)=(\beta,0)$]. As the qubit-detector coupling
strength is reduced (middle row in Fig.~2), the measurement bases
start to deviate from the charge basis, and they develop some
statistical spread. This region could be called the
intermediate-coupling regime. Deep in the weak-coupling regime
(bottom row in Fig.~2), the measurement bases are spread over all
the possible directions. The probability distribution of possible
measurement bases will be analyzed in Sec.~IV.D.

When $\beta=0$ (leftmost column in Fig.~2), the measurement basis
is always the charge basis, regardless of the qubit-QPC coupling
strength. This result is natural, since this case is the simple
one with no mixing between the states $\ket{L}$ and $\ket{R}$.
When $\beta=\pi/2$, all the possible measurement bases lie in the
$x$-$y$ plane. The reason behind this result lies in the fact that
this situation is equivalent to one where one makes a large number
of weak measurements in the $x$-$y$ plane with no
Hamiltonian-induced precession. As explained in Appendix C, the
resulting measurement bases can only be in the same plane as the
actually performed measurements, which is the $x$-$y$ plane in
this case.

The fact that the measurement basis is generally unpredictable,
and therefore uncontrollable, is a rather strange phenomenon from
a fundamental point of view. From a practical point of view, one
can wonder whether anything useful can be done with such
measurements that are performed in a stochastically determined
basis. If one absolutely requires a measurement in a given basis,
measurement results in different bases would be less useful. One
could then treat the deviation of the observed measurement basis
from the desired one as an experimental error and deal with it
accordingly. This point will be discussed further in Sec.~IV.E.

\subsection{Information acquisition rate}

\begin{figure}[h]
\includegraphics[width=7.0cm]{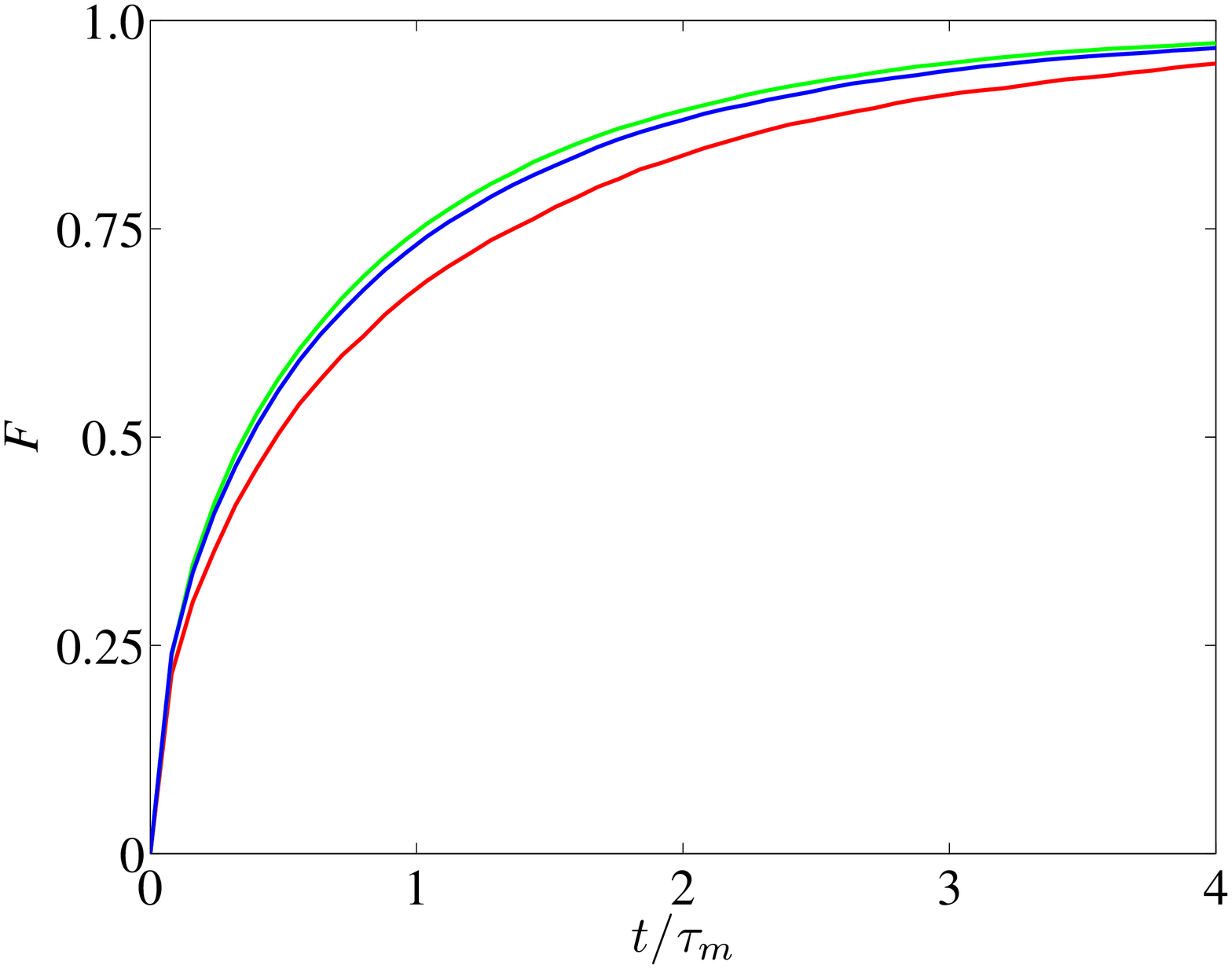}
\caption{(color online) The measurement fidelity as a function of
measurement duration for three different values of the angle
$\beta$ between the charge basis and the energy eigenbasis:
$\beta=0$ (red; lowest line), $\pi/4$ (green) and $\pi/2$ (blue).
Here $E\tau_m/(2\pi)=5$, i.e.~deep in the weak-coupling regime.
The fidelity increases from zero to one as the measurement
duration increases, and it is almost independent of the angle
$\beta$.}
\end{figure}

In Fig.~3, we plot the measurement fidelity as a function of
measurement duration for three different values of $\beta$ (i.e.,
the angle between the charge basis and the qubit's energy
eigenbasis), keeping all other parameters fixed. We can see that
the fidelity approaches one for long enough measurement time,
regardless of the value of the angle $\beta$. Furthermore, the
fidelity is almost independent of $\beta$. In fact, and rather
counter-intuitively, it turns out that the fidelity for $\beta\neq
0$ is higher than that for $\beta=0$. The reason behind this
phenomenon is that one can obtain a higher fidelity by performing
weak measurements along different axes rather than along the same
axis, as explained in Appendix C. This result shows that even
though more complicated analysis is needed to extract useful
measurement information when $\beta \neq 0$, the information
acquisition rate is not reduced, and in fact enhanced, by the
Hamiltonian-induced precession.

In the limit when the fidelity reaches one, the matrix
$\hat{U}_{\rm Meas}[I(t:0\rightarrow t_f),\delta \! I,\delta t]$
has the form $\ket{\psi_1}\bra{\psi_1}$. It is straightforward to
verify that any further evolution of the system will not affect
$\hat{U}_{\rm Meas}[I(t:0\rightarrow t_f),\delta \! I,\delta t]$.
In this case, further measurement only changes the final state of
the qubit, which would be given by $\hat{U}_{\rm
Rot}[I(t:0\rightarrow t_f),\delta \! I,\delta t] \ket{\psi_1}$.

It is worth mentioning here that the fidelity for the case
$\beta=0$ is very well fitted by the function
\begin{equation}
F(t) = {\rm erf} \left\{ \sqrt{\frac{t}{2\tau_m}} \right\},
\label{Eq:F_as_function_of_t}
\end{equation}
where ${\rm erf}$ stands for the error function. This result can
be obtained using the following calculation. We take the
probability distribution in
Eq.~(\ref{Eq:Probability_distribution_2}) and we replace $\delta
t$ by $t$. We now identify the portion of this distribution that
corresponds to $I>(\overline{I}_L+\overline{I}_R)/2$ as being the
probability of making a wrong inference about the state of the
qubit. The difference between the probability of making a correct
inference about the state of the qubit and that of making a wrong
inference gives Eq.~(\ref{Eq:F_as_function_of_t}). The fidelity in
Fig.~3 is also relatively well fitted by the function
\begin{equation}
F(t) \approx 1 - \exp \left\{ -\frac{t}{\tau_m} \right\},
\end{equation}
which is an indication that our definition for the measurement
time $\tau_m$ (see Appendix A) is a reasonable one.

\subsection{The information contained in the QPC's output signal}

The QPC produces an output signal that contains fast and large
fluctuations. There are therefore a very large number of possible
output signals that the QPC can display. We now ask the question
of what information is contained in a given QPC output signal. It
turns out that only a small amount of information in the signal
concerns the state of the qubit. As explained above, one can
follow a conceptually straightforward calculation in order to
extract the measurement basis, measurement result and fidelity
from the noisy signal. The signal also contains information about
how the qubit's state is rotated after the effective measurement.
It should be noted here that measurement bases and
post-measurement rotations generally come in all possible
combinations. The measurement basis, result and fidelity on the
one hand and the post-measurement rotation on the other hand can
therefore be seen as two separate pieces of information that are
contained in the QPC's output signal. The rest of the information
contained in the highly noisy signal, which is the vast majority
of information contained in the signal, concerns the QPC itself.
Thus the large fluctuations that one observes provide information
about the state of the QPC in the specific experimental run under
consideration. For purposes of reading out the state of the qubit,
this information is discarded. This situation is most clearly
demonstrated in the case $\beta=0$, where one is only interested
in the time average of the signal, discarding all the fluctuations
around the average.

\subsection{Spectral analysis of the measurement-basis probability distribution}

\begin{figure}[h]
\includegraphics[width=8.0cm]{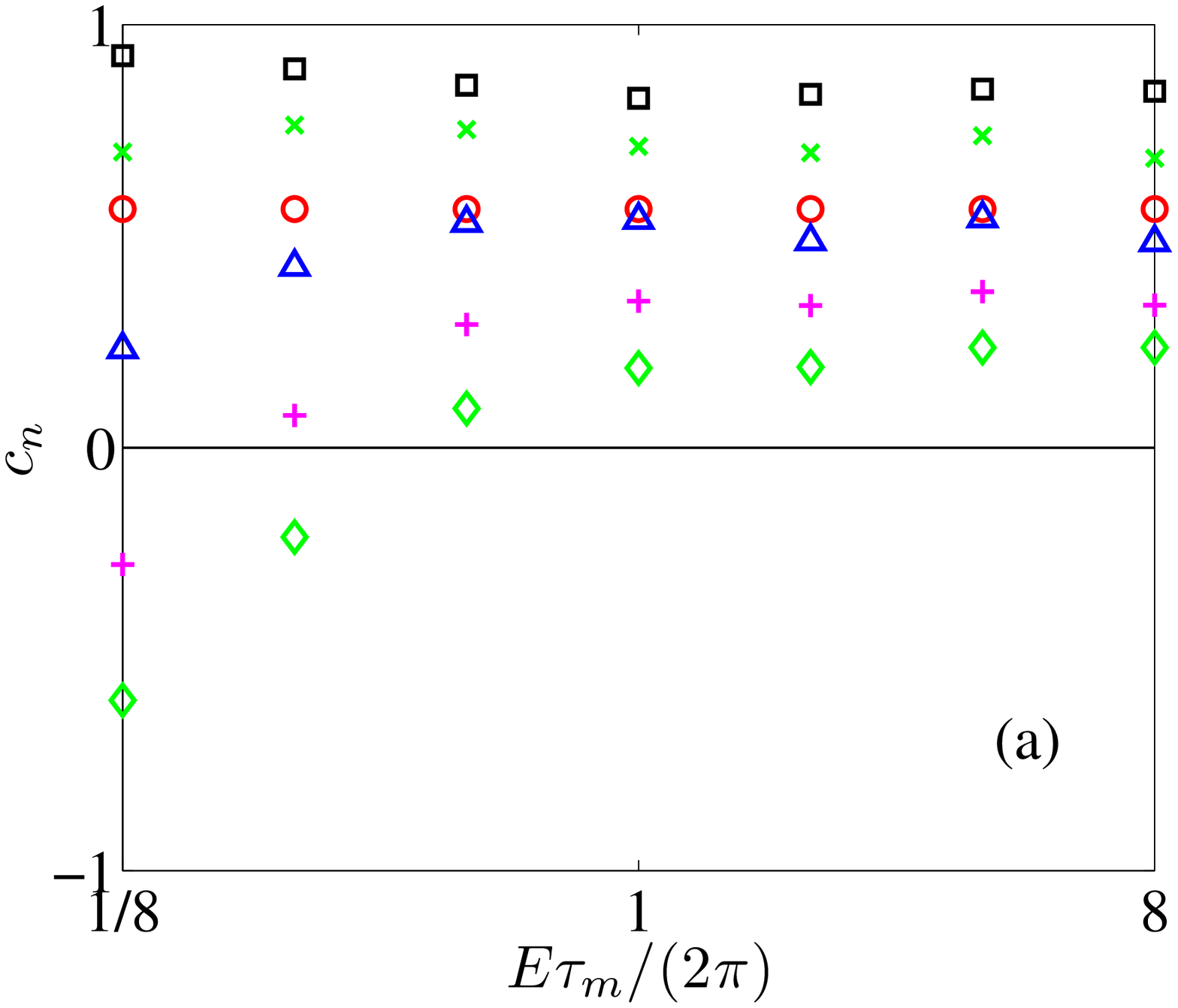}
\includegraphics[width=8.0cm]{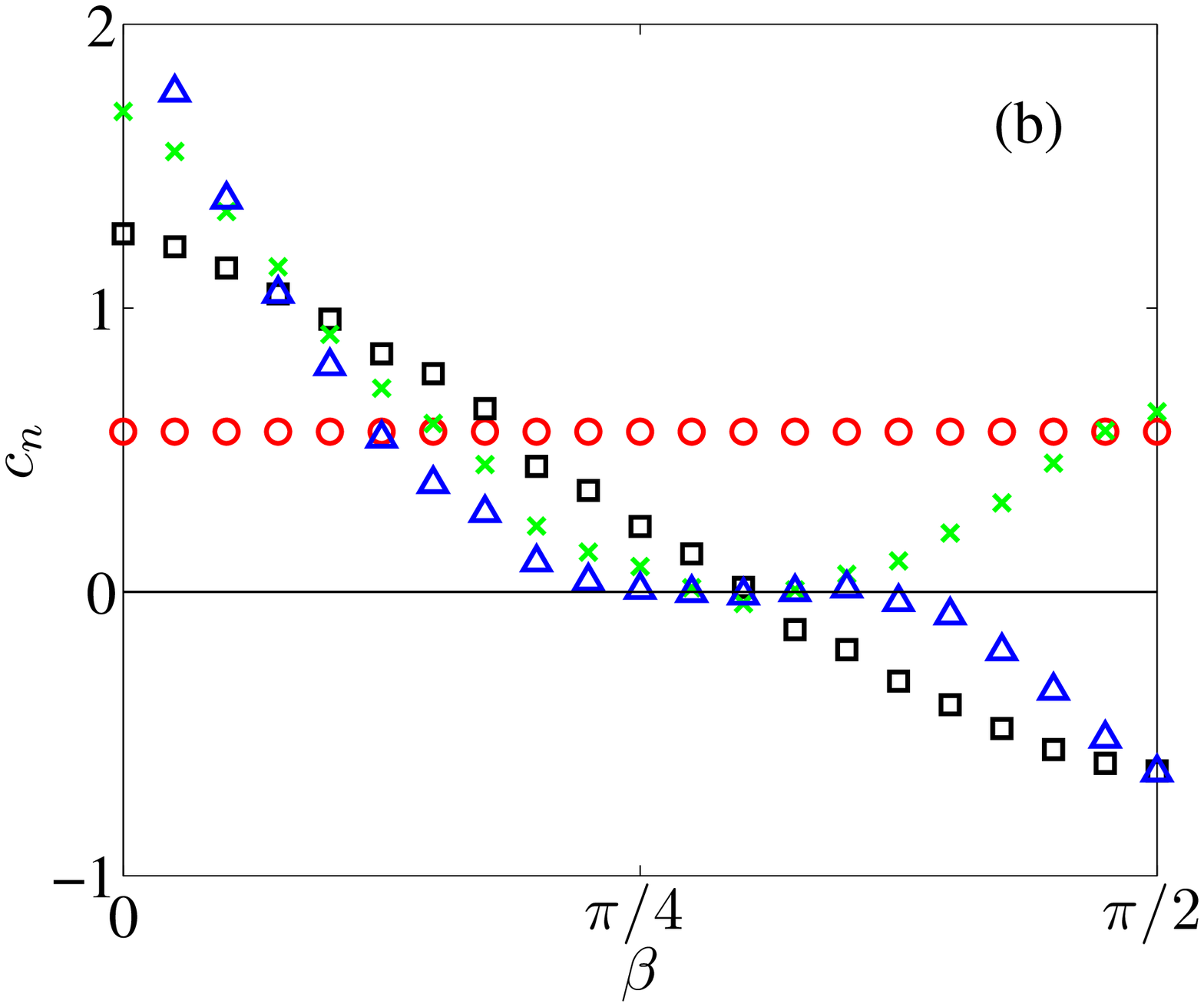}
\caption{(color online) Spectral analysis of the probability
distribution $g(\theta,\phi)$: (a) The spectral components $c_n$
for $n=0$ (red circles), 2 (black squares), 4 (green $\times$
symbols), 6 (blue triangles), 8 (purple + signs) and 10 (green
diamonds) as functions of $E\tau_m/(2\pi)$ for $\beta=\pi/8$. Note
the logarithmic scale on the $x$ axis. As $E\tau_m/(2\pi)$ is
increased, the spectral components clearly reach very stable
values by the time $E\tau_m/(2\pi)$ reaches one. These results
indicate that the value $E\tau_m/(2\pi)=5$ is sufficiently deep in
the weak-coupling regime for purposes of studying the limit $E
\tau_m/(2\pi) \rightarrow \infty$. (b) The spectral components
$c_n$ for $n=0,2,4$ and 6 [same symbol usage as in (a)] as
functions of $\beta$ in the weak-coupling regime [here we use the
value $E\tau_m/(2\pi)=5$]. As $\beta$ is increased, the
probability distribution $g(\theta,\phi)$ gradually changes from
being concentrated at the energy eigenbasis to being spread over
all directions and then to being concentrated on the equator
perpendicular to the energy eigenbasis. There seems to be one
value of $\beta$ [close to $0.3\pi$, which is close to
$\cos^{-1}(1/\sqrt{3})$] where all the spectral components with
$n\neq 0$ vanish, i.e.~$g(\theta,\phi)$ becomes completely
uniform.}
\end{figure}

We now take sets of stochastically determined measurement bases
(i.e., results of the kind presented in Fig.~2) and use them to
extract probability distributions $g(\theta,\phi)$ for the
measurement basis. The probability distributions are normalized to
unity: $\int g(\theta,\phi) d\Omega=1$ with the integral covering
the hemisphere $0\leq\theta\leq\pi/2$, $0\leq\phi\leq 2\pi$. In
particular, we are interested in these probability distributions
in the weak-coupling regime. In this regime, the probability
distributions will be symmetric about the $z$ axis,
i.e.~$g(\theta,\phi)$ will be independent of $\phi$. We use this
fact and write the probability distribution as
\begin{equation}
g(\theta,\phi) = \sum_{n=0}^{\infty} c_n Y_{n,0}(\theta,\phi),
\end{equation}
where $Y_{n,0}(\theta,\phi)$ are the usual spherical harmonics
given by
\begin{equation}
Y_{n,0}(\theta,\phi) = \sqrt{\frac{2n+1}{4\pi}} P_n(\cos\theta),
\end{equation}
and $P_n(x)$ are the Legendre Polynomials. Note that since we only
need to consider the hemisphere $0\leq\theta\leq\pi/2$, we only
need to keep even values of $n$ in the above series. The
coefficients $c_n$ are given by
\begin{equation}
c_n = 2 \int_0^{\pi/2} \!\!\! d\theta \sin \theta \int_{0}^{2\pi}
\!\! d\phi \; g(\theta,\phi) \; Y_{n,0}(\theta,\phi).
\end{equation}
In our calculations, we have a collection of $N$ stochastically
determined measurement bases. The coefficients $c_n$ can be
calculated from these data sets using the formula
\begin{equation}
c_n = \frac{2}{N} \sum_j Y_{n,0}(\theta_j,\phi_j).
\end{equation}
where the index $j$ labels the different data points.

In order to ensure that we are using parameters that are
sufficiently deep in the weak-coupling regime, we start by
performing the above spectrum analysis while varying
$E\tau_m/(2\pi)$. The results of this calculation for
$\beta=\pi/8$ are shown in Fig.~4(a). The coefficients $c_n$ seem
to be almost independent of $E \tau_m/(2\pi)$ when this parameter
is larger than one. From these results we conclude that the point
$E \tau_m/(2\pi)=5$ is sufficiently deep in the weak-coupling
regime for purposes of calculating $g(\theta,\phi)$ in the limit
$E \tau_m/(2\pi) \rightarrow \infty$.

We now calculate $c_n$ (up to $n=10$) in the weak-coupling regime
for different values of $\beta$ and plot the results in Fig.~4(b).
The probability distribution changes gradually as we go from
$\beta=0$, where the measurement basis is always along the $z$
axis (here the charge basis), to $\beta=\pi/2$, where the
measurement basis is always in the $x$-$y$ plane. There is one
point ($\beta\approx 0.3 \pi$) where $g(\theta,\phi)$ is either
approximately or exactly uniform over the entire hemisphere.

\subsection{Short-time dynamics}

We now consider the case where the measurement duration is short
enough that the measurement fidelity is much smaller than one. In
this case the state of the qubit experiences only a small amount
of projective evolution. As a result we can calculate analytically
the total effect of the projection and coherent dynamics.

Dividing the total time interval under consideration into a large
number of infinitesimal intervals, we can write the total
evolution matrix as
\begin{widetext}
\begin{eqnarray}
\hat{U}_{\rm Total} \left[ I(t:0\rightarrow t_f), \delta \! I,
\delta t \right] & \approx & \left[ 1 + \delta p(t_f)
\hat{\sigma}_{\bf n} \right] e^{i \hat{\sigma}_z E\delta t/2}
\cdots e^{i \hat{\sigma}_z E\delta t/2} \left[ 1 + \delta p(\delta
t) \hat{\sigma}_{\bf n} \right] e^{i \hat{\sigma}_z E\delta t/2}
\left[ 1 + \delta p(0) \hat{\sigma}_{\bf n} \right]
\nonumber
\\
& = & e^{i \hat{\sigma}_z E t_f/2} \left[ 1 + \delta p(t_f)
\hat{\sigma}_{{\bf n} (t_f)} \right] \times \cdots \times \left[ 1
+ \delta p(\delta t) \hat{\sigma}_{{\bf n} (\delta t)} \right]
\times \left[ 1 + \delta p(0) \hat{\sigma}_{{\bf n} (0)} \right]
\nonumber
\\
& \approx & e^{i \hat{\sigma}_z E t_f/2} \left[ 1 + \delta p(t_f)
\left\{ \sin\beta \cos(Et_f) \hat{\sigma}_x + \sin\beta \sin(Et_f)
\hat{\sigma}_y + \cos\beta \hat{\sigma}_z \right\} \right] \times
\cdots
\nonumber
\\
& & \hspace{3cm} \times \left[ 1 + \delta p(0) \left\{ \sin\beta
\cos(0) \hat{\sigma}_x + \sin\beta \sin(0) \hat{\sigma}_y +
\cos\beta \hat{\sigma}_z \right\} \right]
\nonumber
\\
& \approx & e^{i \hat{\sigma}_z E t_f/2} \left[ 1 + \int_0^{t_f}
dt \frac{\delta p(t)}{\delta t} \left\{ \sin\beta \cos(Et)
\hat{\sigma}_x + \sin\beta \sin(Et) \hat{\sigma}_y + \cos\beta
\hat{\sigma}_z \right\} \right]
\nonumber
\\
& = & e^{i \hat{\sigma}_z E t_f/2} \left[ 1 + p_x \hat{\sigma}_x +
p_y \hat{\sigma}_y + p_z \hat{\sigma}_z \right],
\end{eqnarray}
\end{widetext}
where we have defined
\begin{eqnarray}
\delta p(t) & = & \frac{ \delta t \left[ I(t) - \overline{I}
\right]}{\tau_m \Delta\overline{I}}
\nonumber
\\
{\bf n}(t) & = & \sin\beta \cos(Et) \hat{x} + \sin\beta \sin(Et)
\hat{y} + \cos\beta \hat{z}
\nonumber
\\
p_x & = & \sin\beta \int_0^{t_f} \frac{I(t) - \overline{I}}{\tau_m
\Delta\overline{I}} \cos(Et) dt
\nonumber
\\
p_y & = & \sin\beta \int_0^{t_f} \frac{I(t) - \overline{I}}{\tau_m
\Delta\overline{I}} \sin(Et) dt
\nonumber
\\
p_z & = & \cos\beta \int_0^{t_f} \frac{I(t) - \overline{I}}{\tau_m
\Delta\overline{I}} dt.
\end{eqnarray}
Taking the total measurement duration $t_f$ to be a multiple of
the qubit's free-precession period, the different measurements
strengths $p_{x,y,z}$ are different Fourier components of the
observed QPC signal. These Fourier components can be thought of as
independent quantities, leading one to think of the net result as
three independent weak measurements, one along each of the three
axes $x$, $y$ and $z$.

We have performed spectral analysis as was done in Sec.~IV.D on
the possible measurement outcomes in the short-time case analyzed
above (we do not show the full results here). In the calculations
$p_x$, $p_y$ and $p_z$ were assumed to be Gaussian-distributed
quantities with root-mean-square values that obey the
relation
\begin{equation}
p_x^{\rm rms} : p_y^{\rm rms} : p_z^{\rm rms} =
\frac{\sin\beta}{\sqrt{2}} : \frac{\sin\beta}{\sqrt{2}} :
\cos\beta.
\label{Eq:rms_ratio_of_p}
\end{equation}
The overall features of the dependence of $c_n$ on $\beta$ are
similar to those seen in Fig.~4(b), but there is no exact
agreement between the two cases. For example, the coefficient
$c_2$ is close to being a straight line all the way from $\beta=0$
to $\beta=\pi/2$ in the present case, as opposed to the curved
path that the black squares follow in Fig.~4(b). Finally, we ask
whether there is a value of $\beta$ at which all three
measurements have the same strength on average. Using
Eq.~(\ref{Eq:rms_ratio_of_p}) it is straightforward to see that
this situation occurs when $\cos\beta=1/\sqrt{3}$, i.e.~$\beta=
0.304\pi$; the point at which all measurement directions are
equally probable in Fig.~4(b). The calculations of this subsection
therefore give us a simple interpretation of the significance of
the point $\beta=0.3\pi$ in the present context.

\subsection{Alternative interpretation: Choosing the measurement
basis independently of the measurement outcome}

The idea that the measurement basis is determined stochastically
and is only specified when the experiment is completed is
conceptually unsettling. We therefore take a different approach in
this section. We now require that in order to have a meaningful
measurement, the measurement basis must be specified by the setup
and not by the stochastically determined measurement outcome.

In the strong-coupling regime, there are no conceptual problems.
In this case, the measurement is almost instantaneous, and the
measurement basis is clearly the charge basis. The measurement
fidelity is almost 100\%, up to errors caused by the finiteness of
the measurement time compared to the qubit-precession period.
These errors can be quantified straightforwardly, as we shall do
shortly.

The weak-coupling regime is the one where the conceptual problems
arise. If we take the case where the distribution of measurement
bases is essentially uniform over all direction (which seems to be
the case when $\beta\approx 0.3\pi$), we again obtain a simple,
yet somewhat surprising, result. By symmetry we can argue that all
directions are equivalent. We can therefore choose any basis as
the measurement basis. Based on this choice, we then divide all
the different possible outcomes (i.e.~the different possible QPC
output signals) into two sets, one of which corresponds to the
measurement result $+1$ and the other of which corresponds to the
result $-1$ in the chosen basis. An important question that arises
in this scenario is what the average measurement fidelity would be
under such `worst-case' conditions of having no preferred
measurement basis. In the following, we shall denote this average
measurement fidelity by the symbol $\overline{F}$ in order to
stress the difference between the results of this subsection and
those of Sec.~IV.B. For the completely uniform case, the average
fidelity $\overline{F}$ can be evaluated as follows: let us take a
spin pointing in the $+z$ direction and assume that we have
specified the $z$ axis as the one defining the measurement basis.
If we measure the spin along an axis that makes an angle $\theta$
with the $z$ axis, we obtain the result $+1$ with probability
$\cos^2(\theta/2)$ and the result $-1$ with probability
$\sin^2(\theta/2)$. By dividing the possible outcomes into two
sets (with all outcomes that correspond to states in the upper
hemisphere being interpreted as $+1$ along the $z$ axis and all
outcomes that correspond to states in the lower hemisphere being
interpreted as $-1$ along the $z$ axis) and taking the difference
between the probability of correctly obtaining the result $+1$ and
the probability of obtaining the wrong result $-1$, we find that
the average fidelity (averaged uniformly over the entire
hemisphere)
\begin{eqnarray}
\overline{F}_{{\rm uniform}, E\tau_m\rightarrow\infty} & = &
\int_0^{\pi/2} {\rm d}\theta \sin\theta \left(
\cos^2\frac{\theta}{2} - \sin^2\frac{\theta}{2} \right) \nonumber
\\
& = & \frac{1}{2}.
\end{eqnarray}
The fidelity in this `worst-case scenario' is therefore 50\%. We
emphasize again that this fidelity is independent of our choice of
measurement basis. This result again contrasts the usual intuition
where the outcomes of measurements performed in one basis provide
no information at all in an orthogonal basis. In the present case,
all measurement bases are equivalent, and one would obtain the
same (average) measurement fidelity in any one of them. This
result fits well with the result that will be explained in
Sec.~IV.G that quantum state tomography can be performed by
repeating the same measurement procedure a large number of times.
This measurement procedure gives the same amount of information in
all different bases.

As explained above, the lowest possible average fidelity is 50\%,
and it occurs when $\beta\approx 0.3\pi$. Two other values of
$\beta$ allow for simple reasoning. The obvious one is the case of
$\beta=0$, where the measurement basis is $\{ \ket{L}, \ket{R} \}$
and the fidelity is 100\% for all coupling strengths. The other
exception occurs when $\beta=\pi/2$. As can be seen from the
rightmost column of Fig.~2, the measurement basis is always in the
$x$-$y$ plane. In the weak-coupling regime, the possible
measurement bases are spread uniformly around the equator. In that
limit, one can choose any axis in the $x$-$y$ plane as defining
the measurement basis. All these bases give the same value for the
average fidelity:
\begin{eqnarray}
\overline{F}_{\beta=\pi/2,E\tau_m\rightarrow\infty} & = &
\frac{1}{2\pi} \int_0^{2\pi} {\rm d}\phi \left|
\cos^2\frac{\theta}{2} - \sin^2\frac{\theta}{2} \right| \nonumber
\\
& = & \frac{2}{\pi}.
\end{eqnarray}

Numerical results for the average fidelity for different values of
$\beta$ in the weak-coupling limit are plotted in Fig.~5. First,
in Fig.~5(a) we plot the average fidelity as a function of the
chosen measurement angle $\theta_{\rm Meas}$ for different values
of $\beta$. We find that for $\beta < 0.3\pi$ choosing the energy
eigenbasis as the measurement basis gives the highest average
fidelity, whereas for $\beta > 0.3 \pi$ choosing a basis that is
orthogonal to the energy eigenbasis gives the highest average
fidelity (note that there is an infinite number of such bases). In
Fig.~5(b) we plot the average fidelity as a function of $\beta$
for three choices of the measurement basis: $\theta_{\rm Meas}=0$,
$\theta_{\rm Meas}=\pi/2$ and the charge basis. It should be noted
here that in the weak-coupling limit there is symmetry about the
$z$ axis. As a result, any axis that is obtained by rotating the
charge basis about the Hamiltonian axis will result in the same
value of the fidelity as the charge basis. Choosing the charge
basis as the measurement basis never results in an average
fidelity that is substantially lower than the maximum possible
value.

In the intermediate coupling regime, one can choose the
measurement basis by maximizing the average fidelity. For example,
for the parameters of the middle row in Fig.~2 one can see that
the measurement basis that would maximize the fidelity deviates
slightly from the charge basis (unless $\beta=0$). One can
understand this result naturally by keeping in mind that the
qubit's state precesses about the Hamiltonian's axis while being
measured. The measurement basis is therefore approximately
obtained by taking the charge basis and rotating it by the
appropriate angle about the qubit Hamiltonian (in the opposite
sense from state precession).

\begin{figure}[h]
\includegraphics[width=8.0cm]{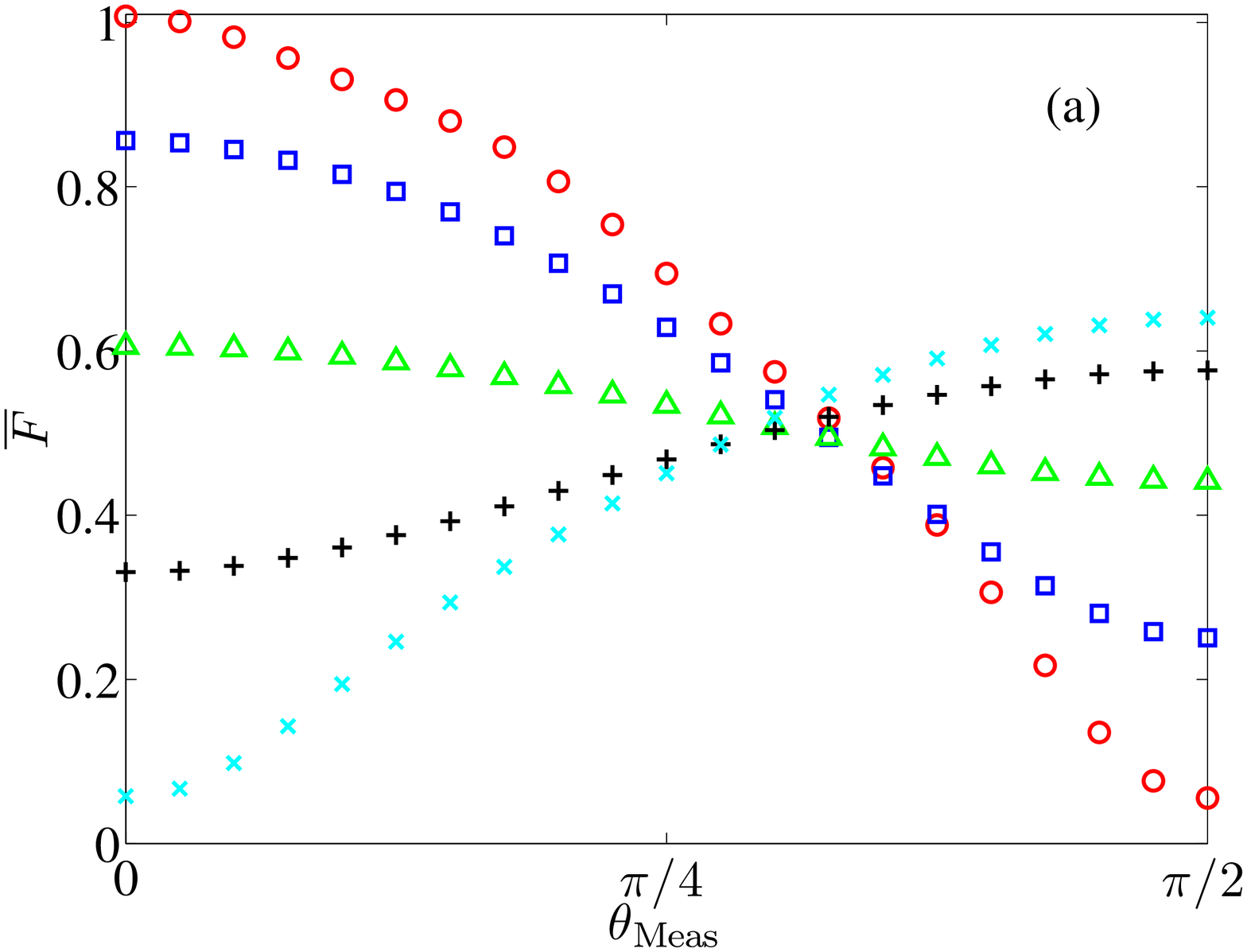}
\includegraphics[width=8.0cm]{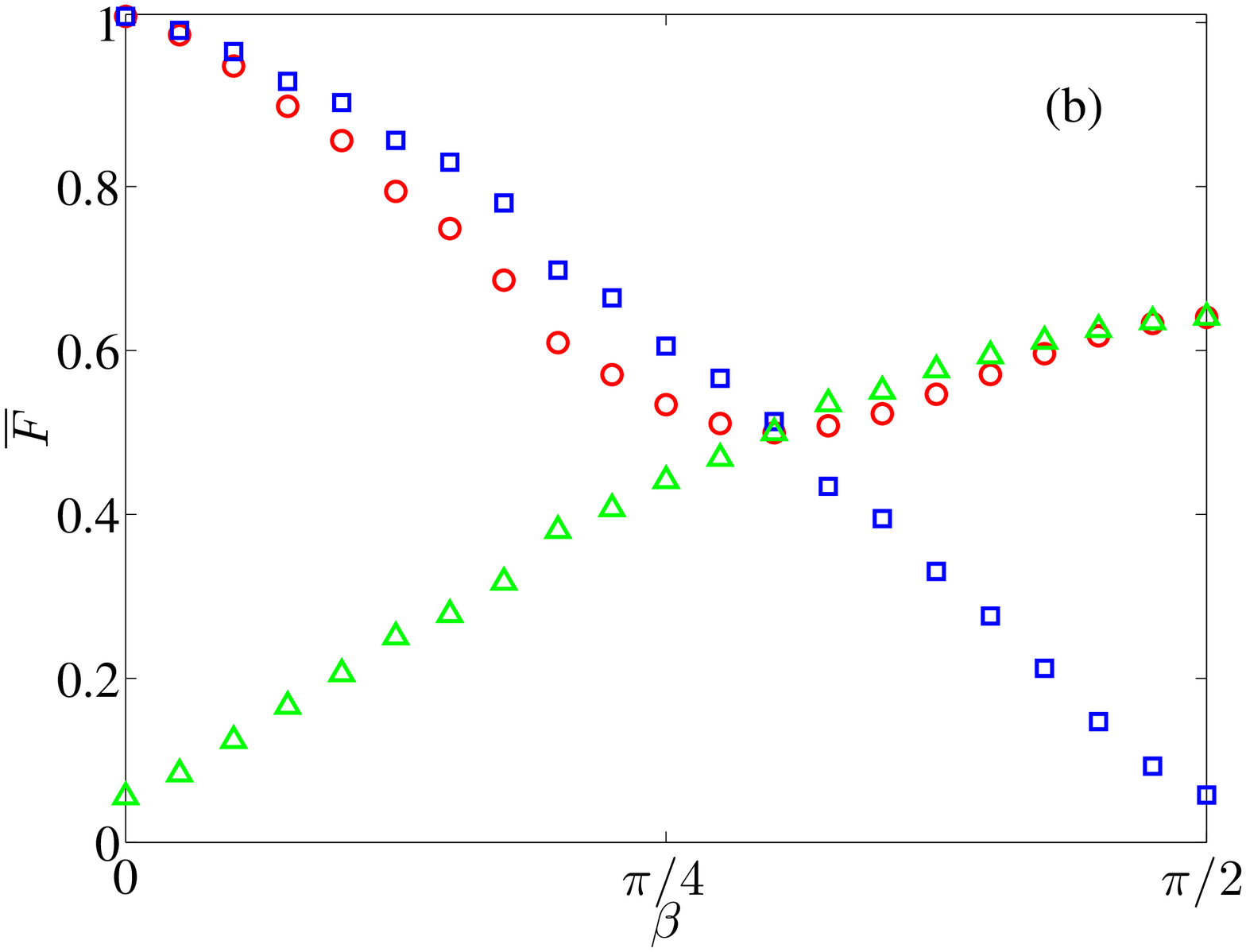}
\caption{(color online) The average fidelity $\overline{F}$ when
defining the measurement basis independently of the specific
(stochastic) outcome in the weak-coupling regime
[$E\tau_m/(2\pi)=5$]: (a) The average fidelity as a function of
the `manually' chosen measurement angle $\theta_{\rm Meas}$ for
five different values of $\beta$: 0 (red circles), $\pi/8$ (blue
squares), $\pi/4$ (green triangles), $3\pi/8$ (black + signs) and
$\pi/2$ (cyan $\times$ symbols). Note that the results are
independent of the choice of the azimuthal angle $\phi_{\rm
Meas}$, which is also chosen manually along with $\theta_{\rm
Meas}$. One can see that for any value of $\beta$ the maximum
fidelity is obtained by choosing either $\theta_{\rm Meas}=0$ (for
$\beta<0.3\pi$) or $\theta_{\rm Meas}=\pi/2$ (for $\beta>0.3\pi$).
For $\beta=0.3$ (not shown) all values of $\theta_{\rm Meas}$ give
essentially the same value for the fidelity, namely 50\%. (b) The
average fidelity as a function of $\beta$ for three different
choices of measurement angle $\theta_{\rm Meas}$: $\theta_{\rm
Meas}=0$ (blue squares), $\theta_{\rm Meas}=\beta$ (red circles)
and $\theta_{\rm Meas}=\pi/2$ (green triangles). The choice
$\theta_{\rm Meas}=\beta$ always gives a near or true maximum in
the fidelity. However, with the exception of a single point around
$\beta=0.3\pi$, the maximum occurs at either $\theta_{\rm Meas}=0$
or $\theta_{\rm Meas}=\pi/2$.}
\end{figure}

\begin{figure}[h]
\includegraphics[width=8.0cm]{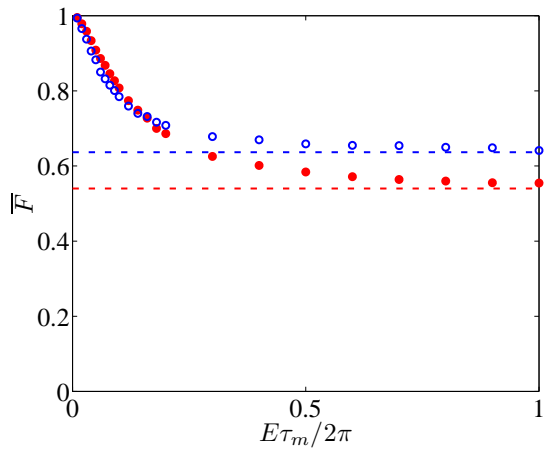}
\caption{(color online) The average fidelity $\overline{F}$ as a
function of $E\tau_m/(2\pi)$ when the measurement basis is decided
(by the experimenter) to be the charge basis. The blue open
circles correspond to the case $\beta=\pi/4$, and the red filled
circles correspond to the case $\beta=\pi/2$. The dashed lines
represent the asymptotic values of the fidelity in the
weak-coupling limit: 0.54 for the case $\beta=\pi/4$ and $2/\pi$
for the case $\beta=\pi/2$.}
\end{figure}

\begin{figure}[h]
\includegraphics[width=8.0cm]{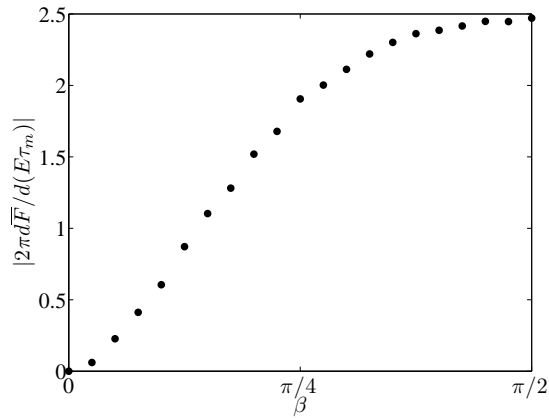}
\caption{The rate of charge-basis fidelity deterioration with
increasing measurement time $|2\pi d\overline{F}/d(E\tau_m)|$ as a
function of $\beta$ in the strong-coupling regime. In other words,
the quantity on the $y$ axis is the absolute value of the slope of
curves similar to those shown in Fig.~6 at $E\tau_m/2\pi=0$. Each
data point is determined by calculating the fidelity
$\overline{F}$ at $E\tau_m/(2\pi)=0.06$ for a given value of
$\beta$ and then taking the ratio $[1-\overline{F}(x)]/x$, where
$x=E\tau_m/(2\pi)$.}
\end{figure}

One could perform a numerical calculation in order to analyze the
behaviour of the maximum-fidelity measurement basis as the
qubit-detector coupling strength is varied. However, here we focus
on a question that seems more experimentally relevant, namely
analyzing the measurement fidelity with the charge basis chosen as
the measurement basis. This fidelity is shown in Fig.~6 for the
two cases $\beta=\pi/4$ and $\beta=\pi/2$. In Fig.~7, we plot the
absolute value of the quantity $2\pi d\overline{F}/d(E\tau_m)$
[i.e., as in the initial slope in Fig.~6] as a function of
$\beta$. This quantity can be used to obtain an estimate for the
measurement errors associated with having a finite measurement
time.

\subsection{Small-current case}

In the above analysis, we have assumed that the QPC's output is
essentially a continuous variable with a large number of possible
values for the current at a given instant. If the QPC's current is
so small that on the timescale of monitoring the output signal
very few electrons can tunnel through the QPC, the physical
picture changes substantially (here we assume that the number of
electrons that have tunneled through the QPC can be measured
accurately). One now has a small number of possible values for the
QPC's output signal. The discretization used in our analysis,
which was done for computational convenience above, is now
provided naturally by the design of the setup. One can then follow
a similar analysis to that of Sec.~III concerning the gradual
progression of the measurement and the evolution of the quantum
state of the qubit. Note that considering the small-current case
only makes sense if the number of electrons that tunnel through
the QPC during a full qubit-precession period is small. Otherwise,
one can always choose between the small- and large-current cases
through the choice of $\delta t$. Clearly, the exact choice of
parameters used in the theoretical analysis cannot affect the
physical results.

An important difference between the cases of small and large
currents is that one expects the probability distributions shown
in Fig.~1(b) to have almost symmetric shapes for the large-current
case, whereas they almost certainly are not symmetric for the
small-current case. This point is most clearly seen by considering
the situation where at most one electron tunnels though the QPC
during the short interval under consideration. The analysis and
results in this case follow closely the analysis and results for a
switching-based detector, studied recently in
Ref.~\cite{AshhabMeasurement}. The fact that, unlike the case of a
switching-based detector, the measurement now continues after the
occurrence of a tunneling event can be incorporated into the
analysis straightforwardly. Our argument concluding that the
choice of the measurement basis is independent of the initial
state of the qubit breaks down. Perhaps more importantly, the
conceptual picture that the measurement basis is chosen first
(according to some probability distribution) and the measurement
result is then obtained in that basis becomes invalid. For
example, it is now possible for a certain state $\ket{\psi}$ to be
a possible result of the data analysis presented above, but not
the state orthogonal to it \cite{AshhabMeasurement}.

It is also worth mentioning in this context the case where the
tunneling process is coherent on the timescale of qubit
precession. In this case the QPC measures the qubit in the energy
eigenbasis, regardless of the angle $\beta$ (except when
$\beta=\pi/2$ where the QPC fails to perform any measurement on
the qubit). For further discussion of this case, see
e.g.~\cite{Makhlin}, and for a discussion of possible advantages
of weak measurement when dealing with multi-qubit circuits, see
e.g.~\cite{AshhabCoupling}.

\subsection{Quantum state tomography}

One example of a procedure where the uncontrollability of the
measurement basis can be harmless is quantum state tomography
(QST). In fact, the original idea of QST included performing
measurements in all possible bases \cite{DAriano}. We have
simulated QST by repeating the measurement procedure a large
number of times, obtaining a set of measurement results (in the
form of pre-measurement qubit states), and then minimizing the
function (see Appendix D)
\begin{equation}
{\cal T}(r,\theta,\phi) = \sum_j \left[ 1 - r \cos
\Omega(\theta,\phi,\theta_j,\phi_j) \right]^2,
\label{Eq:QST_function}
\end{equation}
where $r$, $\theta$ and $\phi$ are the spherical coordinates of a
point in the Bloch sphere; $j$ is an index labelling the different
runs of the experiment, the direction ($\theta_j,\phi_j$) defines
the qubit state obtained in a given measurement, and
$\Omega(\theta,\phi,\theta_j,\phi_j)$ is the angle between the
directions $(\theta,\phi)$ and $(\theta_j,\phi_j)$. We have chosen
several initial states covering the Bloch sphere, and the
tomography procedure always produced correctly the initial state
of the qubit. For the largest value of the qubit-detector coupling
strength that we used [see top row of Fig.~2], the procedure
became unreliable, because the vast majority of the measurements
are performed in one basis.

It is worth pausing here to ask the question of the minimum
requirements for a single measurement setting to provide full QST
information. The qubit's density matrix is characterized by three
parameters. One therefore needs the ensemble of measurements to
produce three independent variables. If the detector has four
possible outcomes (or output signals), one would obtain three
independent probabilities for three of the outcomes (with the
probability of obtaining the fourth outcome determined by
probability normalization). A four-outcome detector is therefore
sufficient, in principle, to perform full QST on a qubit. One can
generalize this argument straightforwardly to the case of larger
systems: One needs a detector with at least as many possible
outcomes as the square of the size of the Hilbert space in order
to perform QST in that Hilbert space (we do not necessarily imply
pure states here).

In Ref.~\cite{AshhabMeasurement} an `all-in-one' measurement
procedure was analyzed, where the measurement results of
identically prepared setups (using a switching-based detector) can
be used to perform full state tomography of the initial qubit
state as well as extract all relevant system parameters. One could
ask whether a similar situation occurs in the qubit-QPC system.
The answer is yes, in principle. However, because the analysis of
Ref.~\cite{AshhabMeasurement} was almost entirely analytic, it was
possible to obtain simple expressions to which the measurement
data would be fitted in order to extract all the available
parameters. In this paper, we have not been able to obtain similar
analytic expressions. As a result, it would require extensive
numerical calculations to compare the observed data with those
that would be expected for all possible sets of parameters and
then find the parameters that produce the best fit. Given the
relatively long computation times required for even the simple
numerical calculations presented in this paper, we have not
attempted to simulate the full parameter-extraction procedure.

\section{Decoherence}

So far we have assumed that there is no decoherence in the system
other than that associated with the measurement-induced
projection. In this section we briefly consider the effect of
additional decoherence channels on the measurement process.

As a first step, one can make the following observations: the
measurement has a characteristic time $\tau_m$, which determines
how long it takes for the experimenter to obtain a high-fidelity
measurement result. Decoherence introduces its own decoherence
timescales, which we collectively denote by $\tau_d$. If $\tau_m
\ll \tau_d$, decoherence should have little effect on the analysis
and results of the previous sections. If, on the other hand,
$\tau_m \gg \tau_d$, decoherence will distort the state of the
qubit before any reliable measurement result is obtained. This
situation renders the measurement almost completely useless. One
should keep in mind, however, that the above statement depends
strongly on the nature of the decoherence channel. For example, if
the energy eigenbasis coincides with the charge basis (i.e., there
is no coherent dynamics mixing the states $\ket{L}$ and $\ket{R}$)
and there is no relaxation between the states $\ket{L}$ and
$\ket{R}$, pure dephasing between these states will not hamper the
measurement, regardless of the dephasing time.

Decoherence can also be introduced to the stochastic master
equation such that Eq.~(\ref{Eq:SME}) turns into:
\begin{widetext}
\begin{equation}
\dot{\rho}_{\rm q} = - i \left[ \hat{H}_{\rm q} , \rho_{\rm q}
\right] + \frac{\xi(t)}{2\sqrt{\tau_m}} \left( \hat{\sigma}_{\bf
n} \rho_{\rm q} + \rho_{\rm q} \hat{\sigma}_{\bf n} - 2
\left\langle \hat{\sigma}_{\bf n} \right\rangle \rho_{\rm q}
\right) + \frac{1}{4\tau_m} \left( \hat{\sigma}_{\bf n} \rho_{\rm
q} \hat{\sigma}_{\bf n} - \rho_{\rm q} \right) + \sum_k \gamma_k
\left( \hat{A}_k \rho_{\rm q} \hat{A}_k^{\dagger} - \frac{1}{2}
\hat{A}_k^{\dagger} \hat{A}_k \rho_{\rm q} - \frac{1}{2} \rho_{\rm
q} \hat{A}_k^{\dagger} \hat{A}_k \right),
\label{Eq:SME_with_decoherence}
\end{equation}
\end{widetext}
where the index $k$ represents the different decoherence channels,
with $\gamma_k$ and $\hat{A}_k$ being the rate and quantum
operator that describe a given decoherence channel.

Modifying Eq.~(\ref{Eq:SME_for_U}) in order to include decoherence
is somewhat trickier. The difficulty lies in the fact that the
evolution of the density matrix is no longer of the form
\begin{equation}
\rho_{\rm q} \rightarrow \hat{U} \rho_{\rm q} \hat{U}^{\dagger},
\end{equation}
ignoring here terms in Eq.~(\ref{Eq:SME}) that were added for
normalization purposes. One way to deal with this new difficulty
is to note that the evolution of the density matrix caused by
decoherence is linear and therefore write the qubit's density
matrix in vector form, i.e.~the transpose of $(\rho_{00},
\rho_{11}, \rho_{01}, \rho_{10})$. With this rearrangement of the
matrix elements, the evolution matrices are now expressed as
$4\times 4$ matrices acting on the density matrix (which now looks
like a vector) from the left:
\begin{equation}
\rho_{\rm q} \rightarrow \hat{\cal U} \rho_{\rm q},
\end{equation}
where we now use the symbol $\hat{\cal U}$ in order to distinguish
the $4 \times 4$ matrices of this section from the $2 \times 2$
matrices that are applied symmetrically on both sides of the
density matrix.

We shall not write down the detailed expressions for the evolution
matrices $\hat{\cal U}$, because they would not add insight to the
present discussion. Instead we point out that deriving a master
equation for these matrices can be done straightforwardly by
rearranging Eq.~(\ref{Eq:SME_with_decoherence}) into vector
format. The differential equation that describes the evolution of
the density matrix $\rho_{\rm q}$ also describes the evolution of
the propagator $\hat{\cal U}_{\rm Total}$, similarly to the fact
that the Schr\"odinger equation can be applied to wave functions
or propagators. We now assume that the total evolution matrix
$\hat{\cal U}_{\rm Total}$ (including the effects of measurement,
Hamiltonian-induced dynamics and decoherence) has been obtained
using such a stochastic equation, and we ask the question of how
to extract measurement information from it. The answer is a
generalized version of the one obtained in the absence of
decoherence. The total evolution matrix can be decomposed into two
matrices: a measurement matrix followed by a trace-preserving
`quantum operation'. This quantum-operation part describes both
unitary evolution and decoherence \cite{Bengtsson}. The
measurement matrix can be used to extract the measurement basis,
result and fidelity, just as was done in the absence of
decoherence.

We now give a simple example illustrating the interplay between
measurement and decoherence. We assume that the Hamiltonian is
diagonal in the charge basis, and therefore there is no coherent
dynamics between the states $\ket{L}$ and $\ket{R}$. A measurement
matrix that corresponds to an outcome that occurs with probability
$P$ for the maximally mixed state and has fidelity $M$ now takes
the form
\begin{equation}
\hat{\cal U}_{\rm Meas}(P,M) = P \left(
\begin{array}{cccc}
1 \pm M & 0 & 0 & 0 \\
0 & 1 \mp M & 0 & 0 \\
0 & 0 & \sqrt{1-M^2} & 0 \\
0 & 0 & 0 & \sqrt{1-M^2} \\
\end{array}
\right),
\end{equation}
A relaxation matrix with relaxation factor $D_r$ (which can be
thought of as decaying from one to zero as time goes by) is given
by
\begin{equation}
\hat{\cal U}_{\rm relaxation}(D_r) = \left(
\begin{array}{cccc}
1 & 1-D_r & 0 & 0 \\
0 & D_r & 0 & 0 \\
0 & 0 & \sqrt{D_r} & 0 \\
0 & 0 & 0 & \sqrt{D_r} \\
\end{array}
\right),
\end{equation}
and a dephasing matrix with dephasing factor $D_p$ is given by
\begin{equation}
\hat{\cal U}_{\rm dephasing}(D_p) = \left(
\begin{array}{cccc}
1 & 0 & 0 & 0 \\
0 & 1 & 0 & 0 \\
0 & 0 & D_p & 0 \\
0 & 0 & 0 & D_p \\
\end{array}
\right).
\end{equation}
If a qubit is measured, with the measurement outcome corresponding
to the matrix $\hat{\cal U}_{\rm Meas}(P,M)$, and the qubit then
relaxes and dephases according to the matrices $\hat{\cal U}_{\rm
relaxation}(D_r)$ and $\hat{\cal U}_{\rm dephasing}(D_p)$, the
total evolution matrix describing this sequence of events is given
by
\begin{widetext}
\begin{eqnarray}
\hat{\cal U}_{\rm Total} & = & \hat{\cal U}_{\rm dephasing}(D_p)
\hat{\cal U}_{\rm relaxation}(D_r) \hat{\cal U}_{\rm Meas}(P,M) \nonumber \\
& = & P \left(
\begin{array}{cccc}
1 \pm M & (1-D_r) (1 \mp M) & 0 & 0 \\
0 & D_r (1 \mp M) & 0 & 0 \\
0 & 0 & D_p \sqrt{D_r} \sqrt{1-M^2} & 0 \\
0 & 0 & 0 & D_p \sqrt{D_r} \sqrt{1-M^2} \\
\end{array}
\right).
\label{Eq:U_meas_relax_dephas}
\end{eqnarray}
This matrix will be used as a reference matrix for the scenario
that we describe below. In particular, we shall use the fact that
the measurement fidelity is given in terms of the matrix elements
in the top left corner:
\begin{equation}
F = \left| \frac{{\cal U}_{\rm Total}^{1,1}-{\cal U}_{\rm
Total}^{1,2}-{\cal U}_{\rm Total}^{2,2}}{{\cal U}_{\rm
Total}^{1,1}+{\cal U}_{\rm Total}^{1,2}+{\cal U}_{\rm
Total}^{2,2}} \right|.
\end{equation}

We now consider a qubit that is constantly being probed by the QPC
while undergoing gradual relaxation and dephasing. The
measurement-induced evolution is now expressed as
\begin{eqnarray}
\hat{U}_M \rho \hat{U}_M^{\dagger} & \propto & \left[ 1 +
\frac{\delta t \left[ I(t) - \overline{I} \right]}{\tau_m
\Delta\overline{I}} \hat{\sigma}_z \right] \left(
\begin{array}{cc}
\rho_{00} & \rho_{01} \\
\rho_{10} & \rho_{11}
\end{array}
\right) \left[ 1 + \frac{\delta t \left[ I(t) - \overline{I}
\right]}{\tau_m
\Delta\overline{I}} \hat{\sigma}_z \right] \nonumber \\
& \rightarrow & \left(
\begin{array}{cccc}
1 + G(t) + \frac{\delta t}{4\tau_m} & 0 & 0 & 0 \\
0 & 1 - G(t) + \frac{\delta t}{4\tau_m} & 0 & 0 \\
0 & 0 & - \frac{\delta t}{4\tau_m} & 0 \\
0 & 0 & 0 & - \frac{\delta t}{4\tau_m} \\
\end{array}
\right)
\left(
\begin{array}{cc}
\rho_{00} \\
\rho_{11} \\
\rho_{01} \\
\rho_{10}
\end{array}
\right),
\end{eqnarray}
where
\begin{equation}
G(t) = \frac{2 \delta t \left[ I(t) - \overline{I} \right]}{\tau_m
\Delta\overline{I}}.
\end{equation}
Using the above form for the evolution matrix and similar ones for
relaxation and dephasing, we arrive at the equation of motion for
(an unnormalized version of) the total evolution matrix:
\begin{equation}
\delta \hat{\cal U}_{\rm Total} = \delta t \left(
\begin{array}{cccc}
G(t) & \gamma_r & 0 & 0 \\
0 & -\gamma_r - G(t) & 0 & 0 \\
0 & 0 & -\frac{\gamma_r}{2}-\gamma_p-\frac{1}{2\tau_m} & 0 \\
0 & 0 & 0 & -\frac{\gamma_r}{2}-\gamma_p-\frac{1}{2\tau_m} \\
\end{array}
\right) \hat{\cal U}_{\rm Total},
\end{equation}
with the initial condition
\begin{equation}
\hat{\cal U}_{\rm Total}(t=0) = \left(
\begin{array}{cccc}
1 & 0 & 0 & 0 \\
0 & 1 & 0 & 0 \\
0 & 0 & 1 & 0 \\
0 & 0 & 0 & 1 \\
\end{array}
\right).
\end{equation}
The solution of the above differential equation is
\begin{equation}
\hat{\cal U}_{\rm Total}(t) = \left(
\begin{array}{cccc}
\tilde{G}(0,t) & \gamma_r \int_0^{t} e^{-\gamma_r t'}
\tilde{G}^{-1}(0,t') \tilde{G}(t',t) dt' & 0 & 0 \\
0 & e^{-\gamma_r t} \tilde{G}^{-1}(0,t) & 0 & 0 \\
0 & 0 & e^{-(\frac{\gamma_r}{2}+\gamma_p+\frac{1}{2\tau_m})t} & 0 \\
0 & 0 & 0 & e^{-(\frac{\gamma_r}{2}+\gamma_p+\frac{1}{2\tau_m})t} \\
\end{array}
\right),
\end{equation}
\end{widetext}
where
\begin{equation}
\tilde{G}(t_1,t_2) = \exp \left\{ \int_{t_1}^{t_2} G(t) dt
\right\}.
\end{equation}
Comparing this matrix with the one in
Eq.~(\ref{Eq:U_meas_relax_dephas}), one can see that the full
dynamics between times 0 and $t$ is equivalent to a measurement
that is followed by relaxation and dephasing operations. With some
straightforward algebra, one can extract the measurement fidelity:
\begin{eqnarray}
F & = & \left| \frac{\int_0^{t} 2 G(t') e^{-\gamma_r t'}
\tilde{G}^{-2}(0,t') dt'}{2-\int_0^{t} 2 G(t') e^{-\gamma_r t'}
\tilde{G}^{-2}(0,t') dt'} \right| \nonumber \\
& = & \left| \frac{\int_{1}^{\rm final} e^{-\gamma_r t'} d\left[
\tilde{G}^{-2}(0,t')\right]}{2+\int_{1}^{\rm final} e^{-\gamma_r
t'} d\left[ \tilde{G}^{-2}(0,t') \right]} \right|.
\label{Eq:Fidelity_with_decoherence}
\end{eqnarray}
Although the above expression looks rather complicated, one
important result can be seen relatively straightforwardly. The
only difference between this expression and the one that would be
obtained in the absence of decoherence is the factor $e^{-\gamma_r
t'}$. This factor represents the rather intuitive fact that the
parts in the QPC signal that are recorded at later times carry
less value for purposes of determining the initial qubit state
than those recorded at earlier times, with an exponentially
decaying function. What is less intuitively obvious is how the
exponential-decay function enters into the expression for the
fidelity, as can be seen from
Eq.~(\ref{Eq:Fidelity_with_decoherence}).

It is worth making a comment here regarding the `value' of the
output signal in the absence of decoherence. One might be tempted
to say that since the fidelity increase slows down as time goes by
(see Fig.~3), the `value' of the QPC's output signal decreases as
time goes by in that case as well. If that were the case, this
decrease in the `value' of the measurement signal would not be
related to decoherence. However, one can see that this is not the
case by considering the simple case $\beta=0$ in the absence of
decoherence. In that case the important quantity is the
time-averaged QPC current throughout the measurement duration,
without any weight function that decreases in time. All parts of
the signal are therefore given equal importance when extracting
the measurement result.

\begin{figure}[h]
\includegraphics[width=8.0cm]{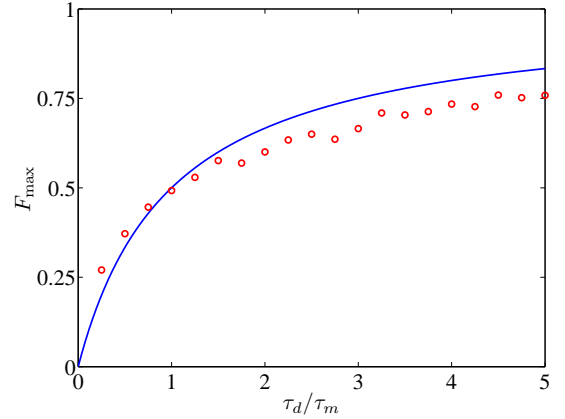}
\caption{The maximum achievable fidelity as a function of the
ratio between the decoherence and measurement times. The red dots
are obtained using numerical simulations, and the solid line is
given by Eq.~(\ref{Eq:Fidelity_with_decoherence_estimate}).}
\end{figure}

In order to calculate the average measurement fidelity in the
presence of decoherence, one must average the measurement fidelity
over all different input states and measurement outcomes. Since
any calculation involving the above expressions seems to be a
daunting task, we use the following hand-waving calculation. In
the absence of decoherence, the measurement fidelity after
duration $t$ can be expressed as
\begin{eqnarray}
F(t) & \approx & 1 - \exp \left\{ -\frac{t}{\tau_m} \right\}
\nonumber
\\
& = & \int_0^t \frac{1}{\tau_m} e^{-t/\tau_m} dt.
\end{eqnarray}
Decoherence reduces the `value' of the later parts of the
measurement record with an exponentially decaying function (with
characteristic time $\tau_d$). An estimate for the average
fidelity in the presence of decoherence can therefore be given in
the form
\begin{eqnarray}
F(t) & \approx & \int_0^t \frac{1}{\tau_m} e^{-t/\tau_m} \times
e^{-t/\tau_d} dt.
\end{eqnarray}
Even if the measurement time is taken to infinity, where we have
$F\rightarrow 1$ in the absence of decoherence, we find that the
fidelity only reaches the value
\begin{equation}
F_{\rm max} = \frac{\tau_d}{\tau_m+\tau_d}.
\label{Eq:Fidelity_with_decoherence_estimate}
\end{equation}
Note, in particular, the simple limits that $F_{\rm max}
\rightarrow 0$ when $\tau_d \ll \tau_m$, and $F_{\rm max}
\rightarrow 1$ when $\tau_d \gg \tau_m$. We have performed
numerical simulations of the measurement process in the presence
of relaxation, and the result are shown in Fig.~8. The results
agree reasonably well with those of the hand-waving argument given
above.

\section{Conclusion}

In conclusion, we have analyzed the question of what information
can be extracted from the output signal of a detector that weakly
probes the state of a qubit when the qubit Hamiltonian induces
oscillations between the two eigenstates of the probed operator.
We have shown that the measurement basis is determined
stochastically every time the measurement is repeated. In the case
of weak qubit-detector coupling, the possible measurement bases
cover all the possible directions. The measurement basis and the
measurement result can both be extracted from the detector's
output signal. We have also shown that the information acquisition
rate is almost independent of the angle $\beta$ between the
direction defining the probed basis and that defining the qubit
Hamiltonian. In other words, given enough time, the detector will
produce a high-fidelity measurement result, regardless of the
value of $\beta$. These results show that, under suitable
conditions and by proper analysis, the detector's ability to
obtain high-fidelity information about the state of the qubit is
not affected by the apparent conflict between the measurement and
coherent-precession dynamics.

In the course of our analysis, we have introduced an equation that
supplements the stochastic master equation for purposes of
extracting measurement information from the observed signal. We
have also introduced a procedure for performing quantum state
tomography that is well suited for the case where the different
measurement bases do not have a simple relation to one another.

We have outlined the generalization of our analysis to the case
where the qubit experiences additional, non-measurement-related
decoherence. In this case, the `value' of the detector's output
signal decreases with the time at which it is recorded. We have
demonstrated this fact by analyzing the case where the measurement
is performed in the presence of relaxation.

We would like to thank D. Burgarth, J. R. Johansson, N. Lambert
and A. J. Leggett for useful discussions. This work was supported
in part by the National Security Agency (NSA), the Laboratory for
Physical Sciences (LPS), the Army Research Office (ARO) and the
National Science Foundation (NSF) grant No.~EIA-0130383. J.Q.Y.
was also supported by the National Basic Research Program of China
grant No. 2009CB929300, the National Natural Science Foundation of
China grant No. 10625416, and the MOST International Collaboration
Program grant No. 2008DFA01930.

\begin{center}
{\bf Appendix A: The QPC-current probability distribution for a
short time interval of size $\delta t$}
\end{center}

In this Appendix, we derive the expression for the probability
distribution $P[I,\delta \! I,\delta t]$ that governs the
stochastically determined values of the QPC current $I$. We use
the discretization parameters $\delta \! I$ and $\delta t$:
$\delta \! I$ defines the size of a finite interval of values for
$I$ that we identify with a single value of $I$, and $\delta t$ is
the size of the time interval under consideration. In the
derivation below, we assume that the qubit is in one of its charge
states (for definiteness $\ket{L}$) and that there is no mixing
between the charge states. We denote the average value of the
current by $\overline{I}_L$ for the state $\ket{L}$ and
$\overline{I}_R$ for the state $\ket{R}$. We are assuming that the
probability distribution can be approximated by a Gaussian
distribution function:
\begin{equation}
P_L[I,\delta \! I,\delta t] \propto \exp \left\{ - \alpha \left(
I-\overline{I}_L \right)^2 \right\}.
\label{Eq:Probability_distribution_1}
\end{equation}

In order to determine how $\delta t$ enters in the expression for
$P_L[I,\delta \! I,\delta t]$ we consider the effect of performing
$N$ measurements in a row. For sufficiently small $\delta t$, the
results of any calculation should not depend on whether we treat
these measurements as $N$ independent measurements or as a single
measurement of length $N\delta t$. Since the values of $I$
obtained in the different measurements are independent, their
average will have a standard deviation $\sigma/\sqrt{N}$, where
$\sigma$ is the standard deviation of $I$ for an interval of
length $\delta t$. In other words, the standard deviation in $I$
must be proportional to $1/\sqrt{\delta t}$. Since the coefficient
$\alpha$ in Eq.~(\ref{Eq:Probability_distribution_1}) is related
to the standard deviation $\sigma$ by $\sigma=1/\sqrt{2\alpha}$,
we find that $\alpha$ must be proportional to $\delta t$:
$\alpha=\tilde{\alpha} \delta t$.

We now introduce the measurement time $\tau_m$ by requiring that
if $\delta t$ were to reach $\tau_m$, the standard deviations (or
in other words, the widths) of the probability distributions
$P_L[I,\delta \! I,\delta t]$ and $P_R[I,\delta \! I,\delta t]$
will be equal to half the distance between their center points:
$1/\sqrt{2\tilde{\alpha} \tau_m} =
|\overline{I}_R-\overline{I}_L|/2$. Note that this definition does
not necessarily agree with other definitions of the measurement
time that appear in the literature. Assuming for definiteness that
$\overline{I}_R>\overline{I}_L$ and defining
$\Delta\overline{I}=\overline{I}_R-\overline{I}_L$, we find that
$\tilde{\alpha} = 2 / \left( \tau_m
\left[\Delta\overline{I}\right]^2 \right)$. Using the
normalization $\sum_{I} P_L[I,\delta \! I,\delta t] = \int
P_L[I,\delta \! I,\delta t] dI/(\delta \! I)=1$, we find that
\begin{equation}
P_L[I,\delta \! I,\delta t] = \sqrt{\frac{2\delta t \left( \delta
\! I \right)^2}{\pi\tau_m \left( \Delta \overline{I} \right)^2}}
\exp \left\{ - \frac{2\delta t \left( I-\overline{I}_L
\right)^2}{\tau_m \left( \Delta\overline{I} \right)^2} \right\}.
\label{Eq:Probability_distribution_2}
\end{equation}
This expression is used for deriving the matrices $U_M[I,\delta \!
I,\delta t]$ in the main text of the paper.

\begin{center}
{\bf Appendix B: The relation between a given output signal and
its `opposite'}
\end{center}

In this appendix we would like to establish the relation between
the measurement results of a certain signal and its `opposite'. If
a signal is given by $I_1(t)$, with the time $t$ running from 0 to
$t_f$, the opposite signal is given by
$I_2(t)=\overline{I}-I_1(t)$. In other words, the opposite signal
is obtained by taking the mirror image of the signal about the
central current value $\overline{I}$. We start by stating the
relation: if
\begin{widetext}
\begin{equation}
\hat{U}_{\rm Total}[I_1(t:0\rightarrow t_f),\delta \! I,\delta t]
= \hat{U}_{\rm Rot}[I_1(t:0\rightarrow t_f),\delta \! I,\delta t]
\times \hat{U}_{\rm Meas}[I_1(t:0\rightarrow t_f),\delta \!
I,\delta t],
\end{equation}
with
\begin{equation}
\hat{U}_{\rm Meas}[I_1(t:0\rightarrow t_f),\delta \! I,\delta t] =
\sqrt{P_1} \ket{\psi_1}\bra{\psi_1} + \sqrt{P_2}
\ket{\psi_2}\bra{\psi_2},
\end{equation}
then
\begin{equation}
\hat{U}_{\rm Total}[I_2(t:0\rightarrow t_f),\delta \! I,\delta t]
= \hat{U}_{\rm Rot}[I_2(t:0\rightarrow t_f),\delta \! I,\delta t]
\times \hat{U}_{\rm Meas}[I_2(t:0\rightarrow t_f),\delta \!
I,\delta t],
\end{equation}
with
\begin{equation}
\hat{U}_{\rm Meas}[I_1(t:0\rightarrow t_f),\delta \! I,\delta t] =
\sqrt{P_1} \ket{\psi_2}\bra{\psi_2} + \sqrt{P_2}
\ket{\psi_1}\bra{\psi_1},
\end{equation}
\end{widetext}
and $\hat{U}_{\rm Rot}[I_2(t:0\rightarrow t_f),\delta \! I,\delta
t]=\hat{U}_{\rm Rot}[I_1(t:0\rightarrow t_f),\delta \! I,\delta
t]$.

For $t_f=0$, both $\hat{U}_{\rm Total}[I_1(t:0\rightarrow
t_f),\delta \! I,\delta t]$ and $\hat{U}_{\rm
Total}[I_2(t:0\rightarrow t_f),\delta \! I,\delta t]$ are given by
the unit matrix, which obeys the relation given above. We now
assume that the relation holds for a given value of $t_f$, and we
try to establish that it will continue to hold at an
infinitesimally later time $t_f+\delta t$. The two changes that
can occur during this infinitesimal interval are a
measurement-induced weak projection or a Hamiltonian-induced
unitary transformation. Since $\hat{U}_{\rm
Rot}[I_2(t:0\rightarrow t_f),\delta \! I,\delta t]=\hat{U}_{\rm
Rot}[I_1(t:0\rightarrow t_f),\delta \! I,\delta t]$, application
of $\exp\{-i\hat{H}\delta t\}$ to $\hat{U}_{\rm
Total}[I_1(t:0\rightarrow t_f),\delta \! I,\delta t]$ and
$\hat{U}_{\rm Total}[I_2(t:0\rightarrow t_f),\delta \! I,\delta
t]$ clearly cannot affect the above relation between the total
evolution matrices. Demonstrating that application of the weak
projection also does not affect the relation is less
straightforward. Using Eq.~(\ref{Eq:UM}) and introducing the
(infinitesimal) quantity $q=\delta t
[I_1(t_f)-\overline{I}]/(\tau_m\Delta \overline{I})$, we find that
\begin{widetext}
\begin{eqnarray}
\hat{U}_{\rm Total}[I_1(t:0\rightarrow t_f+\delta t),\delta \!
I,\delta t] & \propto & \left( 1+q\hat{\sigma}_{\bf n} \right)
\times \hat{U}_{\rm Rot}[I_1(t:0\rightarrow t_f),\delta \!
I,\delta t] \times \left( \sqrt{P_1} \ket{\psi_2}\bra{\psi_2} +
\sqrt{P_2} \ket{\psi_1}\bra{\psi_1} \right)
\nonumber
\\
& = & \hat{U}_{\rm Rot}[I_1(t:0\rightarrow t_f),\delta \! I,\delta
t] \times \left( 1+q\hat{\sigma}_{\tilde{\bf n}} \right) \times
\left( \sqrt{P_1} \ket{\psi_2}\bra{\psi_2} + \sqrt{P_2}
\ket{\psi_1}\bra{\psi_1} \right),
\end{eqnarray}
where
\begin{equation}
\hat{\sigma}_{\tilde{\bf n}} = \hat{U}_{\rm
Rot}^{\dagger}[I_1(t:0\rightarrow t_f),\delta \! I,\delta t]
\times \left( 1+q\hat{\sigma}_{\bf n} \right) \times \hat{U}_{\rm
Rot}[I_1(t:0\rightarrow t_f),\delta \! I,\delta t].
\end{equation}
We now introduce the Pauli matrices $\tilde{\sigma}_z=
\ket{\psi_1}\bra{\psi_1} - \ket{\psi_2}\bra{\psi_2}$ and
$\tilde{\sigma}_x$, for which $q \hat{\sigma}_{\tilde{\bf
n}}=q_x\tilde{\sigma}_x + q_z\tilde{\sigma}_z$. Defining
$a_p=(\sqrt{P_1}+\sqrt{P_2})/2$ and
$a_m=(\sqrt{P_1}-\sqrt{P_2})/2$, we find that
\begin{eqnarray}
\hat{U}_{\rm Total}[I_1(t:0\rightarrow t_f+\delta t),\delta \!
I,\delta t] & \propto & \hat{U}_{\rm Rot}[I_1(t:0\rightarrow
t_f),\delta \! I,\delta t] \times \left( 1 + q_x\tilde{\sigma}_x +
q_z\tilde{\sigma}_z \right) \times \left( a_p + a_m
\tilde{\sigma}_z \right)
\nonumber
\\
& = & \hat{U}_{\rm Rot}[I_1(t:0\rightarrow t_f),\delta \! I,\delta
t] \times \left( a_p + a_p q_x \tilde{\sigma}_x + a_p q_z
\tilde{\sigma}_z + a_m \tilde{\sigma}_z - i a_m q_x
\tilde{\sigma}_y + a_m q_z \right)
\nonumber
\\
& = & \hat{U}_{\rm Rot}[I_1(t:0\rightarrow t_f),\delta \! I,\delta
t] \times \exp \{ - i a_m q_x \tilde{\sigma}_y/a_p \}
\nonumber
\\
& & \hspace{2cm} \times \left( [a_p+a_m q_z] + \left[ a_p q_x -
\frac{a_m^2 q_x}{a_p} \right] \tilde{\sigma}_x + [a_m + a_p q_z]
\tilde{\sigma}_z \right) + O(q^2)
\nonumber
\\
& = & \hat{U}_{\rm Rot}[I_1(t:0\rightarrow t_f+\delta t),\delta \!
I,\delta t] \times \hat{U}_{\rm Meas}[I_1(t:0\rightarrow
t_f+\delta t),\delta \! I,\delta t],
\end{eqnarray}
where
\begin{eqnarray}
\hat{U}_{\rm Rot}[I_1(t:0\rightarrow t_f+\delta t),\delta \!
I,\delta t] & = & \hat{U}_{\rm Rot}[I_1(t:0\rightarrow t_f),\delta
\! I,\delta t] \times \exp \{ - i a_m q_x \tilde{\sigma}_y/a_p \}
\nonumber
\\
\hat{U}_{\rm Meas}[I_1(t:0\rightarrow t_f+\delta t),\delta \!
I,\delta t] & = & a'_p + a'_{m,x} \tilde{\sigma}_x + a'_{m,z}
\tilde{\sigma}_z
\nonumber
\\
a'_p & = & a_p + a_m q_z
\nonumber
\\
a'_{m,x} & = & a_p q_x - \frac{a_m^2 q_x}{a_p}
\nonumber
\\
a'_{m,z} & = & a_m + a_p q_z.
\end{eqnarray}
\end{widetext}
We can now carry out the same derivation as above for the
`opposite' signal. The only changes are that we replace $q$ by
$-q$ (therefore replacing $q_x$ by $-q_x$ and $q_z$ by $-q_z$) and
reverse the positions of $P_1$ and $P_2$ (this latter change
changes $a_m$ into $-a_m$ and leaves $a_p$ unchanged). These
changes leave the infinitesimal rotation angle $a_m q_x/a_p$
unchanged, and we therefore find that the relation $\hat{U}_{\rm
Rot}[I_2(t:0\rightarrow t_f+\delta t),\delta \! I,\delta
t]=\hat{U}_{\rm Rot}[I_1(t:0\rightarrow t_f+\delta t),\delta \!
I,\delta t]$ holds. Similarly $a'_p$ is unaffected by the above
changes. The quantities $a'_{m,x}$ and $a'_{m,z}$, on the other
hand, change their signs. By looking at the expressions for the
$a_p$ and $a_m$ in terms of the $P_1$ and $P_2$ above, we can see
that reversing the sign of $a_m$ reverses the positions of the
symbols $P_1$ and $P_2$ in the matrix $\hat{U}_{\rm Meas}$, which
is exactly the difference between $\hat{U}_{\rm
Meas}[I_1(t:0\rightarrow t_f+\delta t),\delta \! I,\delta t]$ and
$\hat{U}_{\rm Meas}[I_2(t:0\rightarrow t_f+\delta t),\delta \!
I,\delta t]$ that was stated at the beginning of this appendix. We
have thus established the relation between the matrices
$\hat{U}_{\rm Total}[I_1(t:0\rightarrow t_f),\delta \! I,\delta
t]$ and $\hat{U}_{\rm Total}[I_2(t:0\rightarrow t_f),\delta \!
I,\delta t]$.

\begin{center}
{\bf Appendix C: The measurement fidelity for two consecutive weak
measurements}
\end{center}

In this Appendix, we show that the measurement fidelity for two
misaligned weak measurements can be higher than that obtained when
the measurement axes of the two measurements are aligned with each
other \cite{Wei}.

With no loss of generality, we take the measurement axis of the
first measurement to be the $z$ axis (i.e., the basis $\left\{
\ket{0},\ket{1}\right\}$). We assume that the measurement produces
one of two possible outcomes. The measurement matrices can
therefore be expressed as
\begin{eqnarray}
\hat{U}_{1,1} & = & \left(
\begin{array}{cc}
\sqrt{R_1 \left(1+\epsilon_1\right)} & 0 \\
0 & \sqrt{R_1 \left(1-\epsilon_1\right)}
\end{array}
\right) \nonumber
\\
\hat{U}_{1,2} & = & \left(
\begin{array}{cc}
\sqrt{1-R_1 \left(1+\epsilon_1\right)} & 0 \\
0 & \sqrt{1-R_1 \left(1-\epsilon_1\right)}
\end{array}
\right),
\end{eqnarray}
where we have neglected any coherent component in the measurement
matrix, as explained in the main text. The above matrices satisfy
the basic requirement that $\hat{U}_{1,1}^{\dagger}\hat{U}_{1,1} +
\hat{U}_{1,2}^{\dagger}\hat{U}_{1,2} = 1$. The second measurement
is performed in the basis $\left\{\cos(\theta/2) \ket{0} +
\sin(\theta/2) \ket{1},\sin(\theta/2) \ket{0} - \cos(\theta/2)
\ket{1} \right\}$. The corresponding measurement matrices are
given by
\begin{widetext}
\begin{eqnarray}
\hat{U}_{2,1} & = & \sqrt{R_2 \left(1+\epsilon_2\right)} \left(
\begin{array}{cc} \cos^2\frac{\theta}{2} & \sin\frac{\theta}{2}
\cos\frac{\theta}{2} \\
\sin\frac{\theta}{2} \cos\frac{\theta}{2} & \sin^2\frac{\theta}{2}
\end{array}
\right) + \sqrt{R_2 \left(1-\epsilon_2\right)} \left(
\begin{array}{cc} \sin^2\frac{\theta}{2} & -\sin\frac{\theta}{2}
\cos\frac{\theta}{2} \\
-\sin\frac{\theta}{2} \cos\frac{\theta}{2} &
\cos^2\frac{\theta}{2}
\end{array}
\right) \nonumber
\\
\hat{U}_{2,2} & = & \sqrt{1 - R_2 \left(1+\epsilon_2\right)}
\left(
\begin{array}{cc} \cos^2\frac{\theta}{2} & \sin\frac{\theta}{2}
\cos\frac{\theta}{2} \\
\sin\frac{\theta}{2} \cos\frac{\theta}{2} & \sin^2\frac{\theta}{2}
\end{array}
\right) + \sqrt{1 - R_2 \left(1-\epsilon_2\right)} \left(
\begin{array}{cc} \sin^2\frac{\theta}{2} & -\sin\frac{\theta}{2}
\cos\frac{\theta}{2} \\
-\sin\frac{\theta}{2} \cos\frac{\theta}{2} &
\cos^2\frac{\theta}{2}
\end{array}
\right).
\end{eqnarray}

The combined measurement has four possible measurement matrices,
each one corresponding to one of the four possible outcomes. For
example, for the outcome identified with the index 1 in both
measurements, the measurement matrix is given by
$\hat{U}_{2,1}\hat{U}_{1,1}$. In order to obtain the measurement
fidelity for this outcome, we need to consider the matrix
\begin{eqnarray}
\hat{U}_{1,1}^{\dagger} \hat{U}_{2,1}^{\dagger} \hat{U}_{2,1}
\hat{U}_{1,1} & = & \left(
\begin{array}{cc}
\sqrt{R_1 \left(1+\epsilon_1\right)} & 0 \\
0 & \sqrt{R_1 \left(1-\epsilon_1\right)}
\end{array}
\right) \left(
\begin{array}{cc}
R_2 + \epsilon_2 \cos\theta &
\frac{R_2 \epsilon_2 \sin\theta}{2} \\
\frac{R_2 \epsilon_2 \sin\theta}{2} & R_2 - \epsilon_2 \cos\theta
\end{array}
\right) \left(
\begin{array}{cc}
\sqrt{R_1 \left(1+\epsilon_1\right)} & 0 \\
0 & \sqrt{R_1 \left(1-\epsilon_1\right)}
\end{array}
\right).
\label{Eq:U1U2U2U1}
\end{eqnarray}
\end{widetext}
If we denote the eigenvalues of this matrix by $P_1$ and $P_2$
(with $P_1>P_2$), the measurement fidelity for this outcome is
given by $(P_1-P_2)/(P_1+P_2)$. For purposes of calculating the
overall fidelity of the measurement sequence, we take the
maximally mixed state,
\begin{equation}
\rho_{\rm max. \ mixed} = \left(
\begin{array}{cc}
\frac{1}{2} & 0 \\
0 & \frac{1}{2}
\end{array}
\right)
\end{equation}
as the initial state. With this initial state, the probability
that the above outcome is obtained in a given experimental run is
given by $(P_1+P_2)/2$. Thus, the overall fidelity (taking into
account all four possible outcomes) will be given by the sum of
four terms, each of which is given by $(P_1-P_2)/2$ for one of the
possible outcomes. For Eq.~(\ref{Eq:U1U2U2U1}) straightforward
algebra gives the result that
\begin{equation}
\frac{P_1-P_2}{2} = R_1 R_2 \sqrt{(1+\epsilon_1 \epsilon_2
\cos\theta)^2 - (1-\epsilon_1^2) (1-\epsilon_2^2)}.
\end{equation}
Similar expressions can be obtained for the three other possible
outcomes, resulting in the overall fidelity
\begin{eqnarray}
F & = & R_1 R_2 \sqrt{(1+\epsilon_1 \epsilon_2
\cos\theta)^2 - (1-\epsilon_1^2) (1-\epsilon_2^2)} + \nonumber \\
& & R_1' R_2 \sqrt{(1+\epsilon_1' \epsilon_2 \cos\theta)^2 -
(1-\epsilon_1'^2) (1-\epsilon_2^2)} + \nonumber \\
& & R_1 R_2' \sqrt{(1+\epsilon_1
\epsilon_2' \cos\theta)^2 - (1-\epsilon_1^2) (1-\epsilon_2'^2)} + \nonumber \\
& & R_1' R_2' \sqrt{(1+\epsilon_1' \epsilon_2' \cos\theta)^2 -
(1-\epsilon_1'^2) (1-\epsilon_2'^2)},
\end{eqnarray}
where $R_j'=1-R_j$ and $\epsilon_j'=-\epsilon_jR_j/(1-R_j)$.

The measurement fidelity can have its maximum at any point between
zero and $\pi$ (with the proper choice of the parameters $R_j$ and
$\epsilon_j$; we have verified this statement numerically). For
the special case $R_1=R_2=1/2$ and $\epsilon_1=\epsilon_2\equiv
\epsilon>0$, the overall measurement fidelity is given by
\begin{eqnarray}
F & = & \frac{1}{2} \Bigg( \sqrt{(1+\epsilon^2
\cos\theta)^2 - (1-\epsilon^2)^2} + \nonumber \\
& & \hspace{0.7cm} \sqrt{(1-\epsilon^2 \cos\theta)^2 -
(1-\epsilon^2)^2} \Bigg).
\end{eqnarray}
Differentiating this expression twice shows that it has a maximum
at $\cos\theta=0$. In fact, in the limit of weak measurement,
i.e.~$\epsilon\ll 1$, one can see easily that the above expression
can be approximated by
\begin{equation}
F \approx \epsilon \left( \left|\cos\frac{\theta}{2}\right| +
\left|\sin\frac{\theta}{2}\right| \right),
\end{equation}
which has minima at $\theta=0$ and $\theta=\pi$ ($F=\epsilon$) and
a maximum at $\theta=\pi/2$ ($F=\sqrt{2}\epsilon$). The relative
difference between the minimum and maximum fidelities is
approximately 30\%.

The result that the fidelity of two weak measurement can be
enhanced when they are not aligned with one another offers some
explanation for the result of Sec.~IV that the fidelity increases
more rapidly when the qubit Hamiltonian and the probed operator do
not commute. The qubit Hamiltonian causes a rotation in the
qubit's state while the state is being measured along a fixed
axis. This situation is, in some sense, equivalent to one where a
number of different qubit operators are sequentially probed.

A final comment that is worth making here regards the possible
bases of the combined measurement. As can be seen from the
analysis of this appendix, when the two performed measurements
correspond to directions in the $x$-$z$ plane, all the resulting
matrices will be real. As a result, all the possible effective
measurements will be in the $x$-$z$ plane as well (note that a
projector along the $y$ axis would be complex). In other words,
when successive measurements are made in one plane, they can only
result in effective measurements made in the same plane. This
result explains why in the special case $\beta=\pi/2$ all the
possible measurement bases lay in the $x$-$y$ plane; successive
measurements along the $x$ axis separated by state precession
about the $z$ axis are equivalent to a sequence of measurements
that are all performed in the $x$-$y$ plane.

\begin{center}
{\bf Appendix D: Quantum state tomography}
\end{center}

In this Appendix, we derive the function that was used to perform
quantum state tomography in Sec.~IV.

We start with the objective of performing quantum state tomography
using the available measurement data, which we assume has been
obtained using a (generally weakly coupled) QPC as explained in
the main text. We note that since the measurements in the
different experimental runs are generally performed in different
bases, we look for a procedure that treats all the measurements on
equal footing and does not have a preferred measurement basis, or
set of bases. One obvious approach to follow is to construct a
function that assigns a penalty for any deviation by the `guess'
density matrix $\rho_{\rm guess}$ from a given experimental result
(which takes the form of a measured, pure state). The guess
density matrix is characterized by the spherical coordinates
$r_{\rm guess}$, $\theta_{\rm guess}$ and $\phi_{\rm guess}$,
whereas the measured state is characterized by only the angles in
spherical coordinates $\theta_j$ and $\phi_j$ (the index $j$
denotes the different experimental runs). The function to be
minimized for purposes of state tomography will therefore have the
form
\begin{equation}
{\cal T}(r_{\rm guess},\theta_{\rm guess},\phi_{\rm guess}) =
\sum_j f(r_{\rm guess},\theta_{\rm guess},\phi_{\rm
guess},\theta_j,\phi_j).
\end{equation}

In order to proceed further, we now consider what is done in the
standard version of state tomography where the average values
$\avg{\hat{\sigma}_x}$, $\avg{\hat{\sigma}_y}$ and
$\avg{\hat{\sigma}_z}$ are measured and the density matrix $\rho =
(1 + \avg{\hat{\sigma}_x} \hat{\sigma}_x + \avg{\hat{\sigma}_y}
\hat{\sigma}_y + \avg{\hat{\sigma}_z} \hat{\sigma}_z)/2$ is
inferred from the measurement data. If we take the measurement
data for $\hat{\sigma}_z$, we know that we want the tomography
procedure to produce the result $r \cos\theta =
(N_{+1}-N_{-1})/(N_{+1}+N_{-1})$, where $N_{+1}$ and $N_{-1}$ are
the numbers of times that the measurement results $+1$ and $-1$
were obtained, respectively. We can therefore see that the
function $f(r_{\rm guess},\theta_{\rm guess},\phi_{\rm
guess},\theta_j=0,\phi_j)$ will only be a function of
$\avg{\hat{\sigma}_{z,\rm guess}} = r_{\rm guess} \cos\theta_{\rm
guess}$ (or, in other words, the projection of the vector
representing the guess density matrix along the measurement axis).
Taking this measurement data, we find that the function ${\cal T}$
has the form:
\begin{eqnarray}
{\cal T}(\avg{\hat{\sigma}_{z,\rm guess}},N_{+1},N_{-1}) & = &
N_{+1} f(\avg{\hat{\sigma}_{z,\rm guess}}) \nonumber \\ & & +
N_{-1} f(-\avg{\hat{\sigma}_{z,\rm guess}}).
\end{eqnarray}
In order to minimize the function ${\cal T}$, we take its
derivative with respect to $\avg{\hat{\sigma}_{z,\rm guess}}$:
\begin{eqnarray}
\frac{{\rm d} {\cal T}(\avg{\hat{\sigma}_{z,\rm
guess}},N_{+1},N_{-1})}{{\rm d}\avg{\hat{\sigma}_{z,\rm guess}}} &
= & N_{+1} f'(\avg{\hat{\sigma}_{z,\rm guess}}) \nonumber \\ & & -
N_{-1} f'(-\avg{\hat{\sigma}_{z,\rm guess}}).
\end{eqnarray}

If we now use the Taylor expansion
\begin{equation}
f'(\avg{\hat{\sigma}_{z,\rm guess}}) = \sum_{n=0}^{\infty} c_n
\avg{\hat{\sigma}_{z,\rm guess}}^n,
\end{equation}
we find that
\begin{widetext}
\begin{eqnarray}
\frac{{\rm d} {\cal T}(\avg{\hat{\sigma}_{z,\rm
guess}},N_{+1},N_{-1})}{{\rm d}\avg{\hat{\sigma}_{z,\rm guess}}} &
= & \sum_{n=0,2,4,...} \left\{ ( N_{+1} - N_{-1} ) c_n
\avg{\hat{\sigma}_{z,\rm guess}}^n + ( N_{+1} + N_{-1} ) c_{n+1}
\avg{\hat{\sigma}_{z,\rm guess}}^{n+1} \right\} \nonumber \\
& = & ( N_{+1} + N_{-1} ) \sum_{n=0,2,4,...} \left\{ c_n
\avg{\hat{\sigma}_{z,\rm correct}} + c_{n+1}
\avg{\hat{\sigma}_{z,\rm guess}} \right\} \avg{\hat{\sigma}_{z,\rm
guess}}^n,
\end{eqnarray}
\end{widetext}
where $\avg{\hat{\sigma}_{z,\rm correct}} =
(N_{+1}-N_{-1})/(N_{+1}+N_{-1})$. In order to ensure that the
function ${\cal T}$ has a minimum at $\avg{\hat{\sigma}_{z,\rm
guess}} = \avg{\hat{\sigma}_{z,\rm correct}}$, we must choose
$c_n=-c_{n+1}$ for every even number $n$. The simplest choice is
$c_0=-c_1=-1/2$ and $c_2=c_3=...=0$. This choice gives the
function
\begin{equation}
{\cal T}(r_{\rm guess},\theta_{\rm guess},\phi_{\rm guess}) =
\sum_j \left[ 1 - \avg{\hat{\sigma}_{z,\rm guess}} \right]^2,
\end{equation}
which, when generalized to the case where the measurement axis is
not necessarily the $z$ axis, gives Eq.~(\ref{Eq:QST_function}).

The detailed analysis of the different possible choices of $c_n$
in the function ${\cal T}$, and the stability and reliability of
these different choices, is beyond the scope of this paper.


\begin{thebibliography}{99}

\bibitem{Zurek} W. H. Zurek, Phys. Today {\bf 44} (10), 36 (1991);
V. Braginsky, and F. Y. Khalili, {\it Quantum Measurement}
(Cambridge University Press, Cambridge, 1995).

\bibitem{Nielsen} M. A. Nielsen and I. L. Chuang, {\it
Quantum Computation and Quantum Information} (Cambridge University
Press, 2000).

\bibitem{Gurvitz} S. A. Gurvitz, Phys. Rev. B {\bf 56}, 15215 (1997);
B. Elattari and S. A. Gurvitz, Phys. Rev. Lett. {\bf 84}, 2047
(2000); S. A. Gurvitz and G. P. Berman Phys. Rev. B {\bf 72},
073303 (2005); T. Gilad and S. A. Gurvitz, Phys. Rev. Lett. {\bf
97}, 116806 (2006).

\bibitem{Korotkov} A. N. Korotkov, Phys. Rev. B {\bf 60}, 5737 (1999);
A. N. Korotkov, Phys. Rev. B {\bf 63}, 115403 (2001); A. N.
Korotkov and D. V. Averin, Phys. Rev. B {\bf 64}, 165310 (2001);
D. V. Averin and E. V. Sukhorukov, Phys. Rev. Lett. {\bf 95},
126803 (2005).

\bibitem{Makhlin} Y. Makhlin, G. Sch\"on, and A. Shnirman, Phys. Rev.
Lett. {\bf 85}, 4578 (2000).

\bibitem{Goan} H.-S. Goan, G. J. Milburn, H. M. Wiseman, and H. B.
Sun, Phys. Rev. B {\bf 63}, 125326 (2001).

\bibitem{Pilgram} S. Pilgram and M. B\"uttiker, Phys. Rev. Lett.
{\bf 89}, 200401 (2002).

\bibitem{Jordan} A. N. Jordan and M. B\"uttiker, Phys. Rev. B {\bf 71},
125333 (2005); Phys. Rev. Lett. {\bf 95}, 220401 (2006).

\bibitem{Johansson} G. Johansson, L. Tornberg, V. S. Shumeiko, and G.
Wendin, J. Phys.: Condensed Matter {\bf 18}, S901 (2006).

\bibitem{Ouyang} S. H. Ouyang, C. H. Lam, and J. Q. You, J. Phys.:
Condensed Matter {\bf 18}, 11551 (2006).

\bibitem{Reuther} G. M. Reuther, D. Zueco, P. H\"anggi, and S.
Kohler, Phys. Rev. Lett. {\bf 102}, 033602 (2009).

\bibitem{Clerk} A. A. Clerk, M. H. Devoret, S. M. Girvin, F. Marquardt,
R. J. Schoelkopf, arXiv:0810.4729.

\bibitem{Romito} A. Romito, Y. Gefen, and Y. M. Blanter, Phys.
Rev. Lett. {\bf 100}, 056801 (2008); V. Shpitalnik, Y. Gefen, and
A. Romito, Phys. Rev. Lett. {\bf 101}, 226802 (2008).

\bibitem{Kurotani} Y. Kurotani, T. Sagawa, and M. Ueda, Phys. Rev.
A {\bf 76}, 022325 (2007); T. Sagawa and M. Ueda, Phys. Rev. A
{\bf 77}, 012313 (2008); H. Wei and Y. V. Nazarov, Phys. Rev. B
{\bf 78}, 045308 (2008); N. Lambert and F. Nori, Phys. Rev. B {\bf
78}, 214302 (2008).

\bibitem{QuantumCapacitance} We shall not consider measurement
methods based on probing the capacitance of the qubit; M. A.
Sillanp\"a\"a, T. Lehtinen, A. Paila, Yu. Makhlin, L. Roschier,
and P. J. Hakonen, Phys. Rev. Lett. {\bf 95}, 206806 (2005);  T.
Duty, G. Johansson, K. Bladh, D. Gunnarsson, C. Wilson, and P.
Delsing, Phys. Rev. Lett. {\bf 95}, 206807 (2005); S. N.
Shevchenko, S. H. W. van der Ploeg, M. Grajcar, E. Il'ichev, A. N.
Omelyanchouk, and H.-G. Meyer, Phys. Rev. B {\bf 78}, 174527
(2008); E. Il'ichev, S. H. W. van der Ploeg, M. Grajcar, and H. G.
Meyer, Quant. Inf. Proc. {\bf 8}, 133 (2009).

\bibitem{Kreisbeck} A new measurement mechanism in this system was
also proposed in C. Kreisbeck, F. J. Kaiser, and S. Kohler,
arXiv:0904.2754.

\bibitem{Jacobs} K. Jacobs and D. A. Steck, Contemporary Physics
{\bf 47}, 279 (2006); see also T. Brun, Am. J. Phys. {\bf 70}, 719
(2002).

\bibitem{AshhabMeasurement} S. Ashhab, J. Q. You, and F. Nori, Phys. Rev. A
{\bf 79}, 032317 (2009).

\bibitem{AshhabCoupling} S. Ashhab, A. O. Niskanen, K. Harrabi,
Y. Nakamura, T. Picot, P. C. de Groot, C. J. P. M. Harmans, J. E.
Mooij, and F. Nori, Phys. Rev. B {\bf 77}, 014510 (2008); L.
Fedichkin, M. Shapiro, and M. I. Dykman, Phys. Rev. A {\bf 80},
012114 (2009).

\bibitem{DAriano} G. M. D'Ariano, M. G. A. Paris, and M. F.
Sacchi, Advances in Imaging and Electron Physics {\bf 128}, 205
(2003).

\bibitem{Wei} For a study on a related subject, see H. Wei and Y. V.
Nazarov, Phys. Rev. B {\bf 78}, 045308 (2008); a somewhat related
experimental study is given in F. Deppe, M. Mariantoni, E. P.
Menzel, S. Saito, K. Kakuyanagi, H. Tanaka, T. Meno, K. Semba, H.
Takayanagi, R. Gross, Phys. Rev. B {\bf 76}, 214503 (2007).

\bibitem{Bengtsson} I. Bentgsson and K. Zyczkowski, {\it
Geometry of quantum states} (Cambridge University Press, 2006).

\end{thebibliography}
\end{document}